\documentclass[12pt]{article}
\usepackage{amsmath}	
\usepackage{amssymb,amscd}
\usepackage{wrapfig}

\usepackage{amsbsy} 
\usepackage{hyperref}

\usepackage{color}
\usepackage{bm}
\usepackage[all]{xy}

\usepackage{theorem}

\newtheorem{theorem}{Theorem}[section]
 \newtheorem{proposition}[theorem]{Proposition}
 \newtheorem{corollary}[theorem]{Corollary}

{\theorembodyfont{\normalfont}
 \newtheorem{definition}[theorem]{Definition}
 \newtheorem{example}{Example}[section]
 
 }

\newcommand{\qed}{\nobreak \ifvmode \relax \else
      \ifdim\lastskip<1.5em \hskip-\lastskip
      \hskip1.5em plus0em minus0.5em \fi \nobreak
      \vrule height0.75em width0.5em depth0.25em\fi}


\newlength{\minitwocolumn}
\newcommand{\qeq}{\begin{equation}}
\newcommand{\eeq}{\end{equation}}
\newcommand{\bea}{\begin{eqnarray*}}
\newcommand{\eea}{\end{eqnarray*}}
\newcommand{\beqa}{\begin{eqnarray}}
\newcommand{\eeqa}{\end{eqnarray}}

\newcommand{\bX}{\boldsymbol{X}}

\def\ba{{\bm{A}}}

\def\bw{{\bm{W}}}

\def\bz{{\bm{Z}}}
\def\bq{{\bm{Q}}}
\def\bp{{\bm{P}}}

\def\bomega{{\bm{\omega}}}

\def\bR{{\mathbb{R}}}

\def\bZ{{\mathbb{Z}}}

\def\bPhi{{\bm{\Phi}}}

\newcommand{\calA}{{\mathcal A}}

\newcommand{\calC}{{\mathcal C}}
\newcommand{\calD}{{\mathcal D}}

\newcommand{\calF}{{\mathcal F}}

\newcommand{\calL}{{\mathcal L}}
\newcommand{\calM}{{\mathcal M}}
\newcommand{\calN}{{\mathcal N}}
\newcommand{\calO}{{\mathcal O}}

\newcommand{\calS}{{\mathcal S}}

\newcommand{\Map}{{\rm Map}}

\newcommand{\sbv}[2]{{\{{{#1},{#2}}\}}}

\newcommand{\nsbv}[2]{{\{{{#1},{#2}}\}_{\calN}}}

\newcommand{\courant}[2]{{[{{#1},{#2}}]_D}}

\newcommand{\rhoa}{{\rho_{(A)}}}

\newcommand{\bracket}[2]{\langle #1,\,#2\rangle}

\newcommand{\inner}[2]{{({{#1},{#2}})}}

\newcommand{\rd}{\mathrm{d}}

\newcommand{\gomega}{{\omega}}
\newcommand{\ggomega}{{\boldsymbol{\omega}}}

\newcommand{\Thetas}{{\Theta_S}}
\newcommand{\Thetaa}{{\Theta_A}}

\newcommand{\Thetaf}{{\Theta_F}}
\newcommand{\Thetab}{{\Theta_b}}

\newcommand{\SactT}{{S_T^{\nabla}}}
\newcommand{\SactG}{{S_G^{\nabla}}}
\newcommand{\SactGF}{{S_{GF}^{\nabla}}}

\newcommand{\baS}{{{}^A S}}

\newcommand{\anablab}{{{}^A \nabla^{bas}}}
\newcommand{\banabla}{{}^A \stackrel{\bullet}{\nabla}}

\newcommand{\omegam}{{\omega_{\calM}}}
\newcommand{\omegan}{{\omega_{\calN}}}

\newcommand{\Thetam}{{\Theta_{\calM}}}

\newcommand{\Thetag}{{\Theta_G}}
\newcommand{\Thetagf}{{\Theta^{\nabla}_{GF}}}

\begin{document}


\baselineskip 0.7cm

\begin{titlepage}
\begin{flushright}
\end{flushright}

\vskip 1.35cm
\begin{center}
{\Large \bf
Gauged AKSZ sigma models
}
\vskip 1.2cm
Noriaki Ikeda
${}^{a, b}$
\footnote{E-mail:\
nikeda@se.ritsumei.ac.jp
}
\vskip 0.4cm

{\it
${}^a$
Department of Mathematical Sciences,
Ritsumeikan University \\
Kusatsu, Shiga 525-8577, Japan \\
}
${}^b$
{\it Research Institute for Mathematical Sciences, 
Kyoto University \\
Kyoto 606-8502, Japan}
\vskip 0.4cm

\today

\vskip 1.5cm

\begin{abstract}
We propose a new class of sigma models that extend AKSZ sigma models, which we refer to as gauged AKSZ sigma models. These models are constructed using QP-manifolds. The compatibility between the AKSZ structure and gauging is characterized by geometric conditions on the target graded manifold. As an example, we explicitly analyze the gauging of higher Courant sigma models in higher dimensions. When fluxes and boundaries are introduced into gauged higher Courant sigma models, generalizations of momentum maps and Hamiltonian $G$-spaces arise as consistency conditions for the models.
\end{abstract}

\end{center}
\end{titlepage}



\setcounter{page}{2}


\rm

\section{Introduction}
\noindent
AKSZ sigma models are sigma models induced by QP-manifolds, i.e., differential graded symplectic manifolds.\cite{Alexandrov:1995kv, Cattaneo:2001ys, Roytenberg:2006qz}
They encompass not only gauge theories with Lie group symmetries, such as three-dimensional Chern--Simons gauge theory and abelian and nonabelian BF theories in arbitrary dimensions, but also sigma models with algebroid gauge symmetries, such as Poisson sigma models \cite{Ikeda:1993aj, Ikeda:1993fh, Schaller:1994es} and Courant sigma models \cite{Ikeda:2002wh, Hofman:2002rv, Roytenberg:2006qz}.
These models have numerous applications in both mathematics and physics.
The AKSZ construction provides a geometric realization of the Batalin--Vilkovisky (BV) 
\cite{BV1, BV2} and Batalin--Fradkin--Vilkovisky (BFV) formalisms \cite{Batalin:1977pb, Batalin:1983pz} for the corresponding physical theories.

In this paper, we consider the \textit{gauging} of general AKSZ sigma models.
\textit{Gauging} means promoting a global symmetry to a local gauge symmetry by introducing the corresponding gauge fields.
From a mathematical viewpoint, this corresponds to generalizing the target space to one equipped with a group or groupoid action in addition to the original QP-manifold structure.
We refer to the resulting models as \textit{gauged AKSZ sigma models}.
Group or groupoid actions can be described by additional Q-manifold structures.
Thus, gauging amounts to considering the compatibility between two Q-manifold structures.
Let $Q_1$ be the homological vector field of the original Q-manifold, and let $Q_2$ be the homological vector field associated with the gauging Q-manifold.
A natural candidate for the total homological vector field of 
the gauged AKSZ sigma model is
$
Q = Q_1 + Q_2.
$
However, $Q$ does not necessarily satisfy the homological condition, i.e., $Q^2 \neq 0$, unless additional geometric conditions are imposed on the target manifold.
This mathematical structure can be understood in terms of an equivariant differential: $Q^2 = 0$ if and only if certain associated ``curvatures'' are flat.
Such generalizations, or equivariant versions, of BV and AKSZ theories have been discussed in \cite{Bonechi:2012kh, Bonechi:2019dqk, Grigoriev:2024ncm}.
Gauged AKSZ sigma models provide one concrete direction of this framework.

In \cite{Ikeda:2023pdr}, we analyzed the gauging of Poisson sigma models and Dirac sigma models \cite{Kotov:2004wz} by Lie algebroids (and Lie groupoids), including Lie algebras (and Lie groups).
The target-space geometry is described by Poisson manifolds and Dirac structures equipped with group or groupoid actions.
We refer to these models as the gauged Poisson sigma model (GPSM) and the gauged Dirac sigma model (GDSM), respectively.
When the GPSM and GDSM are considered on manifolds with boundaries, the boundary conditions are characterized by Hamiltonian Lie algebroids \cite{Blohmann:2018} over Poisson manifolds \cite{Blohmann:2023} and Dirac structures \cite{Ikeda:2023pdr}.
Another approach to gauging the Poisson sigma model 
was proposed in \cite{Zucchini:2008cg}.
Derivation of the Dirac sigma model by gauging has been analyzed in \cite{Salnikov:2013pwa}.

In our preceding paper \cite{Ikeda:2026mdv}, we analyzed the gauging of Courant sigma models.
Various types of gauging, including gauging by Lie algebroids and Courant algebroids, lead to new sigma models with novel geometric structures.
In the presence of boundaries, the resulting target space structures generalize homotopy momentum maps \cite{Callies:2013jbu} and homotopy momentum sections \cite{Hirota:2021isx}.

In this paper, as concrete examples, we analyze the gauging of AKSZ sigma models in $(n+1)$ dimensions based on higher Courant algebroids on $TM \oplus \wedge^{n-1}T^*M$.
AKSZ sigma models gauged by Lie algebroids, which we refer to as gauged higher Courant sigma models (GHCSMs), are constructed using QP-manifolds.
By introducing various fluxes, we deform the GHCSMs and derive consistency conditions on the fluxes.
These conditions can be regarded as generalizations of the flux identities appearing in string theory \cite{Grana:2008yw, Heller:2016abk}.
From these analyses, we obtain new geometries described by QP-manifolds and AKSZ sigma models, which extend known geometries arising in AKSZ sigma models.
When the source manifold has a boundary, the homological condition $Q^2=0$ gives rise to new geometric structures.
In particular, interesting boundary geometries are described by generalizations of momentum maps, including higher dimensional homotopy moment maps \cite{Callies:2013jbu} and homotopy momentum sections \cite{Hirota:2021isx}.
As a result, we obtain generalizations of momentum maps in the setting of higher Courant algebroids.

In this paper, we do not consider group-like objects corresponding to higher-degree Q-manifolds and higher Courant algebroids.
The analysis of global group-like structures in gauged AKSZ sigma models is left for future work.
For $n$-groupoids and their generalizations, see, for example, the references \cite{Getzler, Henriques, Zhu} and the review \cite{Cueca:2025aej}.

Another important problem is quantization.
The path-integral quantization of the Poisson sigma model leads to the Kontsevich formula for deformation quantization of the target Poisson manifold $M$ \cite{Kontsevich:1997vb, Cattaneo:1999fm}.
For general AKSZ sigma models and their gauged versions, quantization is expected to lead to further generalizations and potentially new interesting results.
The analysis of their quantization is also left for future work.

This paper is organized as follows.
In Section 2, we summarize basic facts about 
QP-manifolds and AKSZ sigma models with examples 
induced from Lie and higher Courant algebroids.
In Section 3, we discuss the general theory of gauging of AKSZ sigma models.
In Section 4, we introduce boundaries in gauging AKSZ sigma models.
In Section 5, gauging of higher Courant sigma models is calculated 
as a concrete example.
In Section 6, Fluxes and boundaries are introduced in gauged higher Courant sigma models in Section 5.
In the Appendix, we summarize definitions, formulas and 
local coordinate expressions of Lie algebroids.

\paragraph{Notation}

$i, j, k$, etc are for indices of local coordinates in a target base manifold 
$M$, and its tangent and cotangent bundle.
$a, b, c$, etc are for indices of local coordinates of 
the fiber of a vector bundle $E$ and $A$.
The Einstein summation convention is used throughout, so that summation 
symbols are omitted when an index appears both as an upper and a lower index.

\section{Preliminary: Q-manifolds, QP-manifolds and AKSZ sigma models}
\label{app:qmanifold}
Sigma models called AKSZ sigma models \cite{Alexandrov:1995kv} are reviewed in this section.
For more details, refer to articles, for instance, \cite{Cattaneo:2001ys,  Roytenberg:2006qz, Ikeda:2012pv}.

\subsection{Q-manifolds}
\begin{definition} 
A \textit{graded manifold} $\calM$ is a locally ringed space $(M, \calO_{M})$ 
over an ordinary smooth manifold $M$, whose structure sheaf $\calO_{M}$ is a $\bZ$-graded commutative algebra. The grading is compatible with the supermanifold grading, that is, a variable of even degree is commutative and a variable of odd degree is anticommutative. By definition, the structure sheaf of $M$ is locally isomorphic to $C^{\infty}(U)\otimes S^{\bullet}(V)$, where $U$ is a local chart on $M$, $V$ is a graded vector space, and $S^{\bullet}(V)$ is a free graded commutative ring on $V$.
$\calO_{M}$ is also denoted by $C^{\infty}(\calM)$.
Grading of an element $f \in C^{\infty}(\calM)$
is called degree and denoted by $|f|$.
For two homogeneous elements $f, g \in C^{\infty}(\calM)$, the product is graded commutative, $fg = (-1)^{|f||g|}gf$.
\end{definition} 
There are some recent reviews of graded manifolds \cite{Roytenberg:2006qz, Cattaneo:2010re, Cueca:2025aej} and references therein.
We concentrate on nonnegatively graded manifolds called an N-manifold.
\begin{definition}
If an N-manifold $\calM$ has a vector field $Q$ of degree $+1$ satisfying $Q^2=0$, it is called a \textit{Q-manifold}, or a \textit{differential graded (dg) manifold}.
\end{definition} 
This vector field $Q$ is called a homological vector field.

\begin{example}[Lie algebroids]\label{LAexample}
Let $A$ be a vector bundle over $M$.
$A$ is a Lie algebroid if and only if $A[1]$ is a Q-manifold 
with a homological vector field \cite{Vaintrob}.
For the definition and properties of Lie algebroids, refer to Appendix \ref{app:Liealgebroid}.
For a vector bundle $A$ over a smooth manifold $M$, $A[1]$ is a grade manifold
such that degree of fiber coordinates is shifted by one.
The concrete local expression is as follows.
Let $(x^i, a^a)$ be local coordinates on $A[1]$ of degree $(0, 1)$.
Here $i$ is the index on the base manifold $M$ and $a$ is 
the index of the fiber.
A vector field of degree $+1$ is locally given by
\begin{align}
Q &= \rho^i_a(x) a^a \frac{\partial}{\partial x^i}
- \frac{1}{2} C_{bc}^a(x) a^b a^c \frac{\partial}{\partial a^a},
\label{homologicalQ}
\end{align}
where $\rho^i_a(x)$ and $C^a_{bc}(x)$ are local functions.
We define a bundle map $\rho: A \rightarrow TM$ and the Lie bracket 
$[-,-]: \Gamma(A) \times \Gamma(A) \rightarrow \Gamma(A)$
as
\begin{align}
\rho(e_a) &:= \rho^i_a(x) \partial_i,
\\ 
~[e_a, e_b] &:= C_{ab}^c(x) e_c.
\end{align}
for the basis of sections $e_a \in \Gamma(A)$.
Then, $Q^2=0$ is equivalent that the two operations $(\rho, [-,-])$ 
satisfy the definition of a Lie algebroid on $A$.
all the identities of the anchor map and the Lie bracket in 
the Lie algebroid $A$ are obtained from the $Q^2=0$ condition.
\end{example}

\subsection{QP-manifolds}

\begin{definition} 
If an N-manifold $\calM$ has a graded symplectic form $\gomega$ of degree $n$ 
and a vector field $Q$ of degree $+1$ satisfies $Q^2=0$, 
and $\calL_Q \gomega =0$,
$(\calM, \gomega, Q)$ is called a \textit{QP-manifold} of degree $n$.
\end{definition} 
For a QP-manifold $\calM$, a graded Poisson bracket on $C^\infty ({\cal M})$ 
is defined by
$$    
\{f,g\} 
= (-1)^{|f|+n} \iota_{X_f} \delta g
= (-1)^{|f|+n+1} \iota_{X_f} \iota_{X_g} \gomega,
$$
for $f, g \in C^\infty({\cal M})$,
where the Hamiltonian vector field $X_f$ is given by the equation 
$\iota_{X_f} \omega = - \delta f$. 
In a QP-manifold with $n \neq 0$, there always exists a 
Hamiltonian function $\Theta$ such that $Q = \sbv{\Theta}{-}$.
$Q^2 = 0$ is equivalent to $\sbv{\Theta}{\Theta}=0$.
$\Theta$ is also called a homological function.

\begin{example}[Lie algebroids]\label{LAQP}
Let $A$ be a vector bundle over a smooth manifold $M$.
We consider a $2$-shifted cotangent bundle $\calM= T^*[2]A[1]$.

Take local coordinates on $T^*[2]A[1]$,
$(x^i, z_i, a^a, w_a)$ of degree $(0, 2, 1, 1)$,
where $i= 1, \ldots, \dim(M)$ and $a = 1, \ldots, \mathrm{rank}(A)$.
Since $T^*[2]A[1]$ is a (graded) cotangent bundle, it 
has a canonical graded symplectic form of degree $2$ given by
\begin{align}
\gomega &= \delta x^i \wedge \delta z_i 
+ \delta a^a \wedge \delta w_a.
\label{gsymplecticl}
\end{align}

A QP-manifold structure on $T^*[2]A[1]$ is defined by introducing 
a homological function $\Theta$ of degree three. We consider 
the following function,
\begin{align}
\Theta &= \rho^i_a(x) z_i  a^a + \frac{1}{2} C_{ab}^c(x) a^a a^b w_c.
\label{CAhomologicalfnl}
\end{align}
Imposing the condition $\sbv{\Theta}{\Theta}=0$, a QP-manifold structure on 
$T^*[2]A[1]$ is given.
\begin{theorem}
A QP-manifold with the homological function \eqref{CAhomologicalfnl}
has one to one correspondence to a Lie algebroid $E$.
\end{theorem}
A Lie algebroid structure on $A$ is recovered from a 
QP-manifold $T^*[2]A[1]$ by derived brackets.
In fact, operations of a Lie algebroid are given by
\begin{align}
\sbv{e_1}{e_2} &= -\sbv{\sbv{e_1}{\Theta}}{e_2},
\\
\rho(e)f &= -\sbv{\sbv{e}{\Theta}}{f}.
\end{align}
Here $f \in C^{\infty}(M)$ and $e, e_1, e_2 \in C^{\infty}_1(A[1])$, 
where $C^{\infty}_1(A[1])$ is the space of degree one functions,
which is identified to the space of sections $\Gamma(A)$.
\end{example}

\begin{example}[Courant algebroids]\label{CAQP}
Let $E$ be a vector bundle over a smooth manifold $M$.
Assume an inner product on the fiber $\bracket{-}{-}$.
About the definition of the Courant algebroid, refer to \cite{LWX, Roy01}.
In this paper, a Courant algebroid appears only in this example, 
as a structure on $E$ induced from a QP-manifold of degree $2$, 
$T^*[2]E[1]$.

We consider a $2$-shifted cotangent bundle $\calM= T^*[2]E[1]$.
Take local coordinates on $T^*[2]E[1]$,
$(x^i, z_i, \eta^a)$ of degree $(0, 2, 1)$,
where $i= 1, \ldots, \dim(M)$ and $A= 1, \ldots, \mathrm{rank}(E)$.
Here $E$ and $E^*$ is identified by the inner product $\bracket{-}{-}$.
The local coordinate expression of the inner product is 
$k_{ab} = \bracket{e_a}{e_b}$ for the basis $e_a \in \Gamma(E)$.
Since $T^*[2]E[1]$ is a (graded) cotangent bundle, it 
has a canonical graded symplectic form of degree $2$ given by
\begin{align}
\gomega &= \delta x^i \wedge \delta z_i 
+ \frac{1}{2} \delta \eta^a \wedge \delta (k_{ab} \eta^b).
\label{gsymplectic}
\end{align}

A QP-manifold structure on $T^*[2]E[1]$ is defined by introducing a homological 
function $\Theta$ of degree three. The most general general form is 
\begin{align}
\Theta &= \rho^i_a(x) z_i \eta^a + \frac{1}{3!} H_{ABC}(x) \eta^a \eta^a \eta^a.
\label{CAhomologicalfn}
\end{align}
Imposing the condition $\sbv{\Theta}{\Theta}=0$, a QP-manifold structure on 
$T^*[2]E[1]$ is given.
There are a basic theorem for a QP-manifold of degree two.
\begin{theorem}\cite{Roy01}
A QP-manifold of degree two has one to one correspondence to a Courant algebroid $E$.
\end{theorem}
A Courant algebroid structure on $E$ is recovered from a QP-manifold of 
degree two $T^*[2]E[1]$ by derived brackets.
In fact, operations of a Courant algebroid are given by
\begin{align}
\bracket{e_1}{e_2} &= \sbv{e_1}{e_2},
\\
\courant{e_1}{e_2} &= -\sbv{\sbv{e_1}{\Theta}}{e_2},
\\
\rho(e)f &= -\sbv{\sbv{e}{\Theta}}{f},
\\
\calD f &= \sbv{\Theta}{f}.
\end{align}
Here $f \in C^{\infty}(M)$ and $e, e_1, e_2 \in C^{\infty}_1(E[1])$, 
where $C^{\infty}_1(E[1])$ is the space of degree one functions,
which is identified to $\Gamma(E)$.

On the other hand, for a Courant algebroid $E$,
the homological function \eqref{CAhomologicalfn}
on $T^*[2]E[1]$ is constructed from the anchor map $\rho$
and the Dorfman bracket $\courant{-}{-}$ by
taking $\rho(e_a) = \rho^i_a(x) \partial_i$, 
$\bracket{e_a}{e_b} = k_{ab}$
and $\courant{e_a}{e_b} = H_{abc}(x) k^{cd} e_d$
for the basis $e_a \in \Gamma(E)$.
\end{example}

\begin{example}[Higher Courant algebroids]\label{standardNAQP}
We consider the $n$-shifted cotangent bundle $\calM= T^*[n]T[1]M$.
Higher QP-manifolds and induced higher Dorfman brackets have been
analyzed in \cite{Hagiwara, Wade, BS, Zambon:2010ka, Ikeda:2012pv, BMR}

Take local coordinates on $T^*[n]T[1]M$
$(x^i, z_i, q^i, p_i)$, of degree $(0, n, 1, n-1)$.
Here $x^i$ is a local coordinate on $M$ and $q^i$ is local coordinate 
of the fiber of $T[1]M$. $(z_i, p_i)$ are dual coordinates of the fiber direction $T^*[n]$.
Since $T^*[n]T[1]M$ is a (graded) cotangent bundle, it 
has a canonical graded symplectic form of degree $n$ given by
\begin{align}
\gomega &= \delta x^i \wedge \delta z_i 
+ (-1)^n \delta q^i \wedge \delta p_i.
\label{gsymplectic2}
\end{align}
$\gomega$ induces a graded Poisson bracket $\sbv{-}{-}$ of degree $-n$,
which bracket for canonical conjugates are
\begin{align}
\sbv{x^i}{z_j} &= - \sbv{z_j}{x^i} = \delta^i_j,
\\
\sbv{q^i}{p_j} &= (-1)^n \sbv{p_j}{q^i} = \delta^i_j.
\end{align}
A QP-manifold structure on $T^*[n]T[1]M$ is defined by introducing a homological 
function $\Theta$ of degree $n+1$. 
Here we consider the following homological function,
\begin{align}
\Theta &= z_i q^i + \frac{1}{(n+1)!} H_{i_1\ldots i_{n+1}}(x) q^{i_1} \ldots q^{i_{n+1}},
\label{CAhomologicalfn3}
\end{align}
where $H = \frac{1}{(n+1)!} H_{i_1\ldots i_{n+1}}(x) 
\rd x^{i_1} \wedge \ldots \wedge \rd x^{i_{n+1}}
\in \Omega^{n+1}(M)$
is an  $(n+1)$-form on $M$.
If and only if $H$ is closed,  $\Theta$ is homological, i.e., $\sbv{\Theta}{\Theta}=0$.
Then, $(\omega, \Theta)$ give a QP-manifold structure of degree 
$n$ on $T^*[n]T[1]M$.

Under the condition $\sbv{\Theta}{\Theta}=0$, a QP-manifold structure on 
$T^*[n]T[1]M$ induces a higher algebroid structure on 
$TM \oplus \wedge^{n-1} T^*M$ as follows.

Let us consider a degree $n-1$ function $u^i(x) p_i + \frac{1}{(n-1)!}
\alpha_{i_1 \ldots i_{n-1}}(x) q^{i_1} \ldots q^{i_{n-1}} 
\in C_{n-1}^{\infty}(T^*[n]T[1]M)$. It is identified to
a section $u + \alpha = u^i(x) \partial_i + 
\frac{1}{(n-1)!}
\alpha_{i_1 \ldots i_{n-1}}(x) \rd x^{i_1} \wedge \ldots \wedge 
\rd x^{i_{n-1}} 
\in \Gamma(TM \oplus \wedge^{n-1} T^*M)$.
Therefore, $C^{\infty}_{n-1}(T^*[n]T[1]M) \simeq 
\Gamma(TM \oplus \wedge^{n-1} T^*M)$.
Obviously, the space of degree zero functions is 
$C^{\infty}(M)$, i.e.,
$C^{\infty}_{0}(T^*[n]T[1]M) \simeq  C^{\infty}(M)$.

Derived brackets on $C^{\infty}_{0}(T^*[n]T[1]M) \oplus
C_{n-1}^{\infty}(T^*[n]T[1]M)$
give operations of higher Courant algebroids 
on $TM \oplus \wedge^{n-1} T^*M$.
In fact, operations are given by
\begin{align}
\courant{u + \alpha}{v + \beta} &= 
- \sbv{\sbv{u + \alpha}{\Theta}}{v + \beta} 
= [u, v] + \calL_{u}\beta - \iota_{v} \rd \alpha + \iota_{u}
\iota_{v} H,
\label{ndorfman}
\\
\rho(u + \alpha)f &= - \sbv{\sbv{u + \alpha}{\Theta}}{f} = u  f,
\label{nanchor}
\\
\bracket{u+\alpha}{v+\beta} & = \sbv{u+\alpha}{v+\beta}
= \iota_u \beta + \iota_v \alpha,
\label{ninner}
\end{align}
where $\courant{-}{-}:\Gamma(TM \oplus \wedge^{n-1} T^*M) \times 
\Gamma(TM \oplus \wedge^{n-1} T^*M) \rightarrow 
\Gamma(TM \oplus \wedge^{n-1} T^*M)$ is the Dorfman bracket,
$\rho: E \rightarrow TM$ is the anchor map, 
and 
$\bracket{-}{-}$ is a bilinear inner product on 
$\Gamma(TM \oplus \wedge^{n-1} T^*M)$.
Identities between three opeartions
\eqref{ndorfman}, \eqref{nanchor} and \eqref{ninner} are obtained 
form the homological condition of $\Theta$, $\sbv{\Theta}{\Theta}=0$
and identities of the graded Poisson bracket $\sbv{-}{-}$.

Resulting algebroids with three operations and identities induced from 
$\sbv{\Theta}{\Theta}=0$
are called a higher Courant algebroid on $TM \oplus \wedge^{n-1} T^*M$.
They are also called symplectic Lie $n$-algebroids.
We obtain the following proposition.
\begin{proposition}
A QP-manifold of degree $n$ induces a higher Courant algebroid 
on $TM \oplus \wedge^{n-1}T^*M$.
\end{proposition}

\end{example}

\subsection{AKSZ sigma models}\label{sec:AKSZ}

If a QP-manifold $(\calM, \gomega, \Theta)$ of degree $n$ is given, 
a QP-manifold structure on a mapping space $\Map(T[1]\Sigma, \calM)$
is defined. It is called an AKSZ sigma model.
Here $\Sigma$ is an ordinary smooth manifold.
If $\Sigma$ is in $n+1$-dimensions, the AKSZ sigma model is equivalent to 
BV-formalism of a gauge theory.
If $\Sigma$ is in $n$-dimensions, the AKSZ sigma model is equivalent to 
BFV-formalism. 

Concrete construction for the BV case is as follows.
Let $\Sigma$ be a manifold in $n+1$ dimensions.
We assume a super differential $\rd$ on $T[1]\Sigma$ induced from the de Rham differential on $\Sigma$ and a Berezin integral measure  $\rd \lambda$ on $T[1]\Sigma_{n+1}$.
Let $\bX: T[1]\Sigma \rightarrow M$ be a map for $\Sigma$ and the base manifold 
$M$ of $\calM$, 

Then the graded symplectic form on $\Map(T[1]\Sigma, \calM)$ is
given by 
\begin{align}
\ggomega &= \int_{T[1]\Sigma} \rd \lambda_{n+1} \, \bX^* \gomega.
\label{AKSZsymplectic}
\end{align}
The BV-AKSZ action functional is given by
\begin{align}
S &= \iota_{\rd} \int_{T[1]\Sigma} \rd \lambda_{n+1} \, \bX^* \vartheta 
+ \int_{T[1]\Sigma} \rd \lambda_{n+1} \, \bX^* \Theta,
\label{AKSZaction01}
\end{align}
where $\rd$ is the super derivative on $T[1]\Sigma$
induced from the de Rham differential on $\Sigma$,
$\vartheta$ is a Liouville $1$-form of $\gomega$ such that 
$\omega = - \delta \vartheta$, 
and $\Theta$ is the homological function for the homological vector field $Q$ 
on $\calM$.

The following theorem has been proved. \cite{Alexandrov:1995kv, Cattaneo:2001ys,Roytenberg:2006qz}
\begin{theorem}
A function $S$ in Eq.~\eqref{AKSZaction01} is a homological function,
i.e., satisfies $\sbv{S}{S}=0$, 
with respect to the graded Poisson bracket induced from Eq.~\eqref{AKSZsymplectic}.
\end{theorem}
$\sbv{S}{S}=0$ is nothing but the classical master equation in BV formalism.
A homological vector field for the function $S$ is 
the sum of two differentials on the mapping space $\Map(T[1]\Sigma, \calM)$ 
induced from two differentials on $T[1]\Sigma$ and $\calM$,
$\sbv{S}{-} = ``\rd + Q'' $. In fact, the Euler-Langrange equations 
induced from the action functional \eqref{AKSZaction01} give
\begin{align}
\rd \circ \varphi^* + \varphi^* \circ Q = 0,
\label{EOM}
\end{align}
for $\varphi \in \Map(T[1]\Sigma, \calM)$.
Eq.~\eqref{EOM} shows that the solution $\phi$ is a Q-manifold morphism
between $T[1]\Sigma$ and $\calM$.

\begin{example}\label{standardNAAKSZ}
As an example, we construct the AKSZ sigma model 
with the target space,
an $n$-shifted cotangent bundle $\calM= T^*[n]T[1]M$
in Example \ref{standardNAQP}.
As explained in Example \ref{standardNAQP}, the QP-manifold structure
on $T^*[n]T[1]M$ induces a higher Courant algebroids
on $TM \oplus \wedge^{n-1} T^*M$.
Take local coordinates on $T^*[n]T[1]M$,
$(x^i, z_i, q^i, p_i)$ of degree $(0, n, 1, n-1)$
as in Example \ref{standardNAQP}.

Let $\Sigma$ be an $(n+1)$-dimensional manifold and
let $(\sigma^{\mu}, \theta^{\mu})$ be local coordinates 
on $T[1]\Sigma$ of degree $(0, 1)$. 
The space of the AKSZ sigma model is the mapping space from 
$T[1]\Sigma$ to $T^*[n]T[1]M$, $\Map(T[1]\Sigma, T^*[n]T[1]M)$.
Local coordinates on $\calM$ is mapped to corresponding superfields
on $\Map(T[1]\Sigma, T^*[n]T[1]M)$.
Superfields which are pullbacks of $(x^i, z_i, q^i, p_i)$ 
are denoted by
by bold letters 
$(\bX^i(\sigma, \theta), \bz_i(\sigma, \theta), 
\bq^i(\sigma, \theta), \bp_i(\sigma, \theta))$
of total degree $(0, n, 1, n-1)$.

A superfield $\bPhi$ is expanded by $\theta^{\mu}$ as
\begin{align}
\bPhi^i(\sigma, \theta) &= \sum_{k=0}^{n+1} 
\frac{1}{k!} \theta^{\mu_1} \ldots \theta^{\mu_k} 
\Phi^{(k)i_1\ldots i_k}(\sigma).
\end{align}
Note that $\theta^{\mu} $ is of degree $1$.
Degree zero coefficient functions 
$\Phi^{(k)i_1\ldots i_k}$ of expansions 
correspond to classical fields.
Degree nonzero parts are ghosts and antifields.
The BV action functional is constructed by the formula 
in the AKSZ construction \eqref{AKSZaction01}, which is
\begin{align}
S &= \int_{T[1]\Sigma} \rd \lambda_{n+1}
\left[(-1)^{n+1} \bz_i \rd \bX^i + (-1)^n \bp_i \rd \bq^i
+ \bz_i \bq^i 
\right.
\nonumber \\
& \left. \qquad 
+ \frac{1}{(n+1)!} H_{i_1\ldots i_{n+1}}(\bX) \bq^{i_1} \ldots \bq^{i_{n+1}}
\right],
\label{HCSMAKSZaciton}
\end{align}
where $\rd$ is the super derivative on $T[1]\Sigma$,
$\rd = \theta^{\mu} \tfrac{\partial}{\partial \sigma^{\mu}}$.
The Berezin measure $\rd \lambda_{n+1}$ in $T[1]\Sigma$
is 
$\rd \lambda_{n+1}
= \rd^{n+1} \sigma \rd^{n+1} \theta
= \rd \sigma^{n+1} \ldots \rd \sigma^{1} 
\rd \theta^{n+1} \ldots \rd \theta^{1}$
in the Cartesian coordinate on $T[1]\Sigma$.
In general, the Berezinian is needed. In this paper, we do not discuss in details. Refer to articles and textbooks about supermanifolds and graded manifolds.
Throughout this paper, 
$\rd$ on $T[1]\Sigma$ is always the super derivative in $T[1]\Sigma$.
From the general theory, $S$ satisfies the homological condition 
$\sbv{S}{S}=0$, which is called the classical master equation
in the BV formalism.
We call this sigma model a \textit{higher Courant sigma model}.

Classical fields corresponding to $(\bX^i, \bz_i, \bq^i, \bp_i)$ are
a map from $\Sigma$ to $M$, $X^i: \Sigma \rightarrow M$,
and differential forms $(Z_i, Q^i, P_i)$ on $\Sigma$.
Precisely, they are
$Z_i = 
\frac{1}{n!} \rd \sigma^{\mu_1} 
\wedge \ldots \wedge \rd \sigma^{\mu_{n}} 
Z^{(n)}_{\mu_1 \ldots \mu_{n} i}(\sigma) 
\in \Omega^n(\Sigma, X^*T^*M)$,
$Q^i = \rd \sigma^{\mu} Q^{(1)i}_{\mu}(\sigma) 
\in \Omega^1(\Sigma, X^*T^*[1]M)$,
and 
\\
$P_i =
\frac{1}{(n-1)!} \rd \sigma^{\mu_1} 
\wedge \ldots \wedge \rd \sigma^{\mu_{n-1}} 
P^{(n-1)}_{\mu_1 \ldots \mu_{n-1} i}(\sigma) 
\in \Omega^{n-1}(\Sigma, X^*T[1]M)$.

The classical action functional is
\begin{align}
S_{cl.} 
&= \int_{\Sigma} 
\left[(-1)^{n+1} \inner{Z}{\rd X} + (-1)^{n}\inner{P}{\rd Q}
+ \inner{Z}{Q}
+ H(Q, \ldots, Q)
\right]
\nonumber \\
&= \int_{\Sigma} 
\left[(-1)^{n+1} Z_i \wedge \rd X^i + (-1)^{n}P_i \wedge \rd Q^i
+ Z_i \wedge Q^i 
\right.
\nonumber \\
& \left. \qquad 
+ \frac{1}{(n+1)!} H_{i_1\ldots i_{n+1}}(X) Q^{i_1} \wedge 
\ldots \wedge Q^{i_{n+1}}
\right],
\label{LnASMaction}
\end{align}
where $\rd$ is the ordinary de Rham differential on $\Sigma$.
\end{example}
In remaining sections, we concentrate on AKSZ formulations,
though they always correspond to the BV formalism of corresponding 
classical field theories.

\section{Gauging of AKSZ sigma models}\label{sec:GAKSZ}
In this section, we discuss general structures of gauging 
of AKSZ sigma models.
In our models, gauging is formulated only 
in the target space QP-structure level.

We begin with a target QP-manifold 
$(\calM, \omegam, \Thetam)$ over a smooth manifold $M$.
Gauging is an extension of the graded manifold $\calM$ to 
$\calN$ such that $\calM \subset \calN$.
Here $\calN$ is assumed that there exists a graded symplectic form 
$\omegan$ such that $\omegan|_{\calM} = \omegam$.
In our case, it is assumed that $\calM$ and $\calN$ are the ringed space
with the same base manifold $M$ and the structure sheaf $\calO_{\calM}$
is locally a subsheaf of $\calO_{\calN}$.

An example is as follows.
We consider a vector bundle $E^{\prime}$
and its vector subbundle $E$.
Then, we obtain a pair of grade cotangent bundles
of shifted vector bundles $E[1]$ and $E^{\prime}[1]$,
$\calM = T^*[n]E[1]$ and $\calN = T^*[n]E^{\prime}[1]$.

From the homological function $\Thetam$ on $\calM$, we construct a gauged 
homological function $\Thetag$ on $\calN$ by adding a additional gauging 
term $\Thetaa$.
A naive candidate is $\Thetag = \Thetam + \Thetaa$. 

The first problem of $\Thetag$ is that $\Thetam$ and $\Thetaa$ 
are not necessarily globally well-defined on $\calN$.
It means
that they are not necessarily a global quantity under 
changing of local charts on the structure sheaf on $\calN$, even if 
it is globally well-defined under local coordinate transformations on $\calM$.
$\calN$ has a bigger local coordinate transformations 
since $\calM \subset \calN$.
Our idea is that we 'covariantize' $\Thetam$ and $\Thetaa$
 to $\Thetam^{\nabla}$  and $\Thetaa^{\nabla}$ 
by introducing connections on $\calN$.
We denote the covariantized function by 
\begin{eqnarray}
\Thetag^{\nabla} := \Thetam^{\nabla} + \Thetaa^{\nabla}.
\label{}
\end{eqnarray}
The problem of the choice 
$\Thetag^{\nabla} = \Thetam^{\nabla} + \Thetaa^{\nabla}$ is 
that even if both $\Thetam$ and $\Thetaa$ are homological,
$\Thetam$ is homological with respect to the Poisson 
bracket on $\calM$, however by 
the deformation of $\Thetam$ and $\Thetaa$ to $\Thetam^{\nabla}$
and $\Thetaa^{\nabla}$, 
$\nsbv{\Thetam^{\nabla}}{\Thetam^{\nabla}} \neq 0$ 
and
$\nsbv{\Thetaa^{\nabla}}{\Thetaa^{\nabla}} \neq 0$ 
in general.
$\Thetag^{\nabla}$ does not necessarily satisfy 
the homological condition, i.e., 
in general $\nsbv{\Thetag^{\nabla}}{\Thetag^{\nabla}} \neq 0$
with respect to the graded Poisson bracket induced from the extended 
symplectic form $\omegan$. It means that $Q_G^2 \neq 0$, 
where $Q_G := \sbv{\Thetag^{\nabla}}{-}_{\calN}$ is 
the graded vector field induced from $\Thetag^{\nabla}$.

In this paper, we impose geometric conditions to make $Q_G$
homological, $Q_G^2 = 0$.
In fact, $Q_G$ is a kind of equivariant differential,
which square is proportional to a ``curvature'' $\calF$, 
$Q_G^2 = \calL_{\calF}$, where 
$Q_G$ is decomposed to two terms $Q_G = Q + \iota_{\calF}$ such that $Q^2=0$.
$\calF$ is given by
\begin{eqnarray}
\nsbv{\Thetag^{\nabla}}{\Thetag^{\nabla}} = \calF.
\label{}
\end{eqnarray}
$Q_G$ gives an equivariant differential and $C^{\infty}(\calN)$ becomes
an equivariant complex by decomposing $C^{\infty}(\calN)$ by degree, 
$C^{\infty}(\calN)
= \oplus_k C_k(\calN)$, 
where $C_k(\calN)$ is the space of degree $k$ functions.
If $\calF=0$, $Q_G$ is a coboundary operator and 
$(\calN, \omegan, Q_G)$ is a QP-manifold.
Thus, imposing the flatness condition $\calF=0$, 
an equivariant cohomology $H^{\bullet}(\calN, Q_G)$ is defined.
The structure is in fact a \textit{equivariant QP-structure}.
A homological vector field $Q$ is extended to an equivariant 
vector field such that 
$Q_G = Q + \iota_{\calA} = \sbv{\Theta_G}{-}$
with a 'connection' $\calA$, where $Q_G = \sbv{\Thetag^{\nabla}}{-}$.
$(Q_G)^2 = \iota_{\calF}$ with a 'curvature' $\calF$.
Here $\calF$ is the vector field induced from $\calA$.

The graded symplectic form on the mapping space and
the bulk BV action functional $\SactG = S_{\calM}^{\nabla} + S_A^{\nabla}$
is constructed from the homological function 
$\Thetag^{\nabla} = \Thetam^{\nabla} + \Thetaa^{\nabla}$ 
by the AKSZ construction.
Then $\SactG$ satisfies consistency, 
$\sbv{\SactG}{\SactG} = 0$ if and only if $\calF=0$.
This model is called a {gauged AKSZ sigma model}.

Equivariant versions of BV formalisms and AKSZ theories are analyzed in 
\cite{Bonechi:2019dqk}.
Physical theories with $Q^2 \neq 0$ appear in double field theories
\cite{Chatzistavrakidis:2018ztm, Carow-Watamura:2018iau}.

\section{Gauged AKSZ sigma models with boundary}\label{sec:GAKSZb}

In the gauged AKSZ sigma model on the mapping space 
$\Map(T[1]\Sigma, \calN)$ constructed in Section \ref{sec:GAKSZ},
we consider the base manifold with boundary, i..e, $\Sigma$ such that
$\partial \Sigma \neq \emptyset$, where $\partial \Sigma$ is an $n$-dimensional
manifold.
Let $\iota: \partial \Sigma \rightarrow \Sigma$ be a natural embbeding map.
In this case, boundary conditions are fixed to be consistent with 
bulk QP-manifold structures.
Boundary conditions are determined by introducing boundary terms $S_b$,
which is a boundary integral of the pullback of
a degree $n$ function $\Thetab$ on $\calN$,
\begin{align}
S_b &= \int_{T[1]\partial \Sigma} \rd \lambda_n \, \bX^* \Thetab,
\end{align}
The total BV action is the sum of the bulk action $\SactG$ 
and the boundary term $S_b$.
The bulk action functional is
\begin{align}
\SactG
&= \iota_{\rd} \int_{T[1]\Sigma} \rd \lambda_{n+1} \, \bX^* \vartheta 
+ \int_{T[1]\Sigma} \rd \lambda_{n+1} \, \bX^* \Thetag^{\nabla}.
\label{SactG}
\end{align}
Then, the total action functional is
\begin{align}
\SactT 
&= \SactG + S_b
\nonumber \\ 
&= \iota_{\rd} \int_{T[1]\Sigma} \rd \lambda_{n+1} \, \bX^* \vartheta 
+ \int_{T[1]\Sigma} \rd \lambda_{n+1} \, \bX^* \Thetag^{\nabla}
+ \int_{T[1]\partial \Sigma} \rd \lambda_n \bX^* \Thetab.
\label{SactT}
\end{align}
In general, some geometric conditions on $\calM$ are required in order that
the total BV action functional $\SactT$ satisfies
the homological condition $\sbv{\SactT}{\SactT} = 0$.
Here the symplectic form on the boundary mapping space
$\Map(T[1]\partial \Sigma, \calN)$, 
is given by $\bomega_b = \iota^* \bomega$.
Impose geometric conditions on the target graded manifold $\calM$,
to be $\SactG$ homological, i.e., $\sbv{\SactG}{\SactG} = 0$ 
for $\partial \Sigma = \emptyset$.
Under the conditions, we consider consistent conditions for
boundary conditions.

Assume $\partial \Sigma \neq \emptyset$.
Then, we obtain
$\sbv{\SactT}{\SactT} = 
\sbv{\SactG}{\SactG} + 2 \sbv{\SactG}{S_b} + \sbv{S_b}{S_b}$.
If $\Sigma$ has boundary,  $\sbv{\SactG}{\SactG}$ is not zero any more,
it gives rise to the integral of boundary terms,
\begin{align}
\sbv{\SactG}{\SactG}
&= \iota_{\rd} \int_{T[1]\partial \Sigma} \rd \lambda_{n+1} \, \bX^* \vartheta 
+ \int_{T[1] \partial \Sigma} \rd \lambda_{n} \, \bX^* \Thetag^{\nabla}.
\label{master01}
\end{align}
and 
\begin{align}
\sbv{\SactG}{S_b}
&= \int_{T[1] \partial \Sigma} \rd \lambda_{n} \, \bX^* 
\sbv{\Thetaa^{\nabla}}{\Thetab}.
\label{master02}
\end{align}
$\sbv{S_b}{S_b}$ also give rise to integrals of boundary terms.
We assume $\sbv{S_b}{S_b}=0$,
since an interesting example is given when $\sbv{S_b}{S_b}=0$.
In this case, Eqs.~\eqref{master01} and \eqref{master02} induces that
\begin{align}
\sbv{\SactT}{\SactT} &= \iota_{\rd} \int_{T[1]\partial \Sigma} \rd \lambda_{n+1} \, \bX^* \vartheta 
+ \int_{T[1] \partial \Sigma} \rd \lambda_{n} \, \bX^* \Thetag^{\nabla}
+ 2 \int_{T[1] \partial \Sigma} \rd \lambda_{n} \, \bX^* 
\sbv{\Thetaa^{\nabla}}{\Thetab} 
\label{homologicalboundary}
\end{align}
$\sbv{\SactT}{\SactT} = 0$ is required. 
Then, the first term in Eq.~\eqref{homologicalboundary} must vanish. 
It holds if 
the Liouville one-form  $\vartheta =0$.
It is satisfied if the image of boundary 
$T[1]\partial \Sigma$ is 
in a coisotropic submanifold $\calC$ of the target graded 
manifold $\calM$.
Moreover, the sum of the second and third term in 
Eq.~\eqref{homologicalboundary} must vanish. It gives another condition.
From this analysis, we obtain the following theorem.
\begin{theorem}\label{Generalbc}
Assume $\sbv{S_b}{S_b}=0$.
Then, if and only if 
\begin{align}
(\Thetag^{\nabla} + 2 \sbv{\Thetag^{\nabla}}{\Thetab})|_{\calC} = 0,
\label{boundarymaster}
\end{align}
on a Coisotropic submanifold $\calC \subset \calM$,
then, the gauged AKSZ sigma model with boundary $\SactT$
satisfies the homological condition $\sbv{\SactT}{\SactT}=0$.
\end{theorem}
An important observation is the condition \eqref{boundarymaster}
in Theorem \ref{Generalbc}
is a generalization of the condition of momentum maps.
We have seen it in the papers \cite{Ikeda:2023pdr, Ikeda:2026mdv}
and see it in new examples in later sections.

\section{Gauged higher Courant sigma models}\label{sec:GHCSM}
In this section, we consider a concrete example of 
gauging of higher dimensional AKSZ sigma models,
and analyze problem discussed in the previous section
in concrete examples.

\subsection{Higher Courant sigma models with Lie algebroid gauging}
\label{sec:SCSMLA}
Consider the QP-manifold $T^*[n]T[1]M$ in Example \ref{standardNAQP}.
The corresponding non graded target space is $TM \oplus \wedge^{n-1} T^*M$.
Note that local coordinates on $T^*[n]T[1]M$ are 
$(x^i, z_i, q^i, p_i)$ of degree $(0, n, 1, n-1)$,
where $(x^i, q^i)$ are local coordinates on $T[1]M$ and 
$(z_i, p_i)$ are conjugate coordinates.
Then the AKSZ sigma model is a higher Courant sigma models (HCSM)
in Example \ref{standardNAAKSZ}.
The BV-AKSZ action functional is Eq.~\eqref{HCSMAKSZaciton}.

We consider gauging by Lie algebroids of this model.
Ordinary local gauge symmetries of Lie algebras are special cases of 
Lie algebroids, which is realized as action Lie algebroid structures.

Let $A$ be a Lie algebroid over $M$. For gauging of the HCSM, 
$\calM = T^*[n]T[1]M$ is extended to $\calN = T^*[n](T[1]M \oplus A[1])$.
First, we consider a QP-manifold structure on $T^*[n](T[1]M \oplus A[1])$.
Next, we apply the AKSZ construction to obtain a new AKSZ sigma model.

We take local coordinates on the fiber of $T^*[n]A[1]$, 
$(a^a, w_a)$ of degree $(1, n-1)$
with $a = 1, \ldots, \mathrm{rank}(A)$,
where $a^a$ is a local coordinate of the fiber of $A[1]$ and 
$w_a$ is a canonical conjugate local coordinate on $T^*[n]$.
Since $\calN = T^*[n](T[1]M \oplus A[1])$ is a graded cotangent bundle, 
there exists the canonical graded symplectic form of degree $n$ as,
\begin{align}
\gomega &= 
\delta x^i \wedge \delta z_i 
+ (-1)^n \delta q^i \wedge \delta p_i
+ (-1)^n \delta a^a \wedge \delta w_a,
\label{gsymplectic3}
\end{align}
which satisfies $\gomega_{\calM} = \gomega|_{\calM}$.
From Eq.~\eqref{gsymplectic3}, we obtain graded Poisson brackets 
for canonical quantities,
\begin{align}
\sbv{x^i}{z_j} &= - \sbv{z_j}{x^i} = \delta^i_j,
\\
\sbv{q^i}{p_j} &= (-1)^n \sbv{p_j}{q^i} = \delta^i_j,
\\
\sbv{a^a}{w_b} &= (-1)^n \sbv{w_b}{a^a} = \delta^a_b.
\end{align}
Homological functions in QP-manifolds for higher Courant algebroids 
and Lie algebroids are in 
Eqs.~\eqref{CAhomologicalfn3} and
\eqref{CAhomologicalfnl},
\begin{align}
\Thetas &= \Thetam = z_i q^i + \frac{1}{(n+1)!} H_{i_1\ldots i_{n+1}}(x) q^{i_1} \ldots q^{i_{n+1}},
\label{homological11}
\\
\Thetaa &= \rho^i_a(x) z_i a^a + \frac{1}{2} C_{ab}^c(x) a^a a^b w_c.
\label{homological12}
\end{align}
However they are not invariant under local 
coordinate transformations on $T^*[2](T[1]M \oplus A[1])$.
We see it in the next section \ref{manicovariant}.

If a Lie algebroid $E$ is an action Lie algebroid $A = M \times \mathfrak{g}$,
i.e., symmetries induced from a Lie algebra $\mathfrak{g}$,
we do not need to introduce a nontrivial connection
since homological functions
\eqref{homological11} and \eqref{homological12} 
are already globally well-defined.
After gauging of the action Lie algebroid $M \times \mathfrak{g}$, 
the resulting homological function is invariant under 
local coordinate transformations on 
$T^*[n](T[1]M \oplus A[1]) = T^*[n]T[1]M \oplus T^*[n]\mathfrak{g}[1]$.

\subsection{Local coordinate transformatoins and manifestly 
covariant expressions}\label{manicovariant}
In order to discuss properties of functions $\Thetas$ and $\Thetaa$
under local coordinate transformations,
properties of the grade manifold of degree $n$, $\calM 
= T^*[n](T[1]M \oplus A[1])$ 
are analyzed in local coordinates.

We consider coordinate transformations of local coordinates on 
$T^*[n](T[1]M \oplus A[1])$.
Consider a diffeomorphism transformation on $M$,
$x^{\prime i} = x^{\prime i}(x)$
and a transition function $M^a_b(x)$ in the fiber of the vector bundle $A$, 
where $x^{\prime i}(x)$ is an arbitrary local $C^{\infty}(M)$ function.

Then, transformations of local coordinates on $T^*[n](T[1]M \oplus A[1])$ 
are given by
\begin{align}
& x^{\prime i} = x^{\prime i}(x),
\label{coordinatetransf1}
\\
& q^{\prime i} = \frac{\partial x^{\prime i}}{\partial x^j} q^j,
\label{coordinatetransf2}
\\
& p^{\prime}_i = \frac{\partial x^j}{\partial x^{\prime i}} p_j
\label{coordinatetransf3}
\\
& a^{\prime a} = M^a_b a^b,
\label{coordinatetransf4}
\\
& w^{\prime}_a = M_a^b w_b,
\label{coordinatetransf5}
\\
& 
z^{\prime}_i = \frac{\partial x^j}{\partial x^{\prime i}} z_j 
- \frac{\partial x^k}{\partial x^{\prime l}}
\frac{\partial x^m}{\partial x^{\prime i}}
\frac{\partial^2 x^{\prime l}}{\partial x^m \partial x^j} p_k q^j
- M_c^a \partial_i^{\prime} M^c_b w_a a^c.
\label{coordinatetransf6}
\end{align}
Therefore a graded manifold $T^*[n](T[1]M \oplus A[1])$ is not 
a cotangent bundle in the ordinary sense.

The graded symplectic form is invariant under these local coordinate
transformations \eqref{coordinatetransf1}--\eqref{coordinatetransf6},
\begin{align}
\gomega^{\prime} &= \gomega.
\label{sympinv}
\end{align}
In fact, we can prove Eq.~\eqref{sympinv} as follows.
The Liouville $1$-form $\vartheta$ 
such that $\omega = - \delta \vartheta$ is
\begin{align}
\vartheta & = (-1)^n z_i \delta x^i + (-1)^{n-1} p_i \delta q^i
+ (-1)^{n-1} w_a \delta a^a.
\end{align}
Substituting transformations 
\eqref{coordinatetransf1}--\eqref{coordinatetransf6} directly,
$\vartheta^{\prime} = \vartheta$ is proved.
Invariance of the Liouville $1$-form under local coordinate transformations
induces Eq.~\eqref{sympinv}, i.e., invariance of the symplectic form under local coordinate transformations.

On the other hand, functions $\Thetas$ and $\Thetaa$ are not globally 
well-defined since $z_i$ does not transform covariantly in
Eqs.~\eqref{coordinatetransf1}--\eqref{coordinatetransf6}.

In order to obtain covariant coordinates and global homological functions,
we introduce an affine connection $\nabla_{\mathrm{affine}}: \Gamma(TM) 
\rightarrow \Gamma(TM \otimes T^*M)$
and a vector bundle connection $\nabla_{A}: \Gamma(A) 
\rightarrow \Gamma(A \otimes T^*M)$ on $A$.
We denote the connection $1$-form 
corresponding to the affine connection by $\Gamma_{ij}^k(x) \rd x^i$, 
which satisfies $\nabla_{\mathrm{affine}} \partial_i 
= \Gamma_{ij}^k \rd x^j \otimes \partial_k$,
and the connection $1$-form 
corresponding to the connection on $A$ by $\omega_{ai}^b(x) \rd x^i$, 
which satisfies $\nabla_A e_a = \omega_{ai}^b \rd x^i \otimes e_b$.\footnote{In the paper \cite{Ikeda:2026mdv}, an affine connection has not been introduced since we are interested in consistency conditions with Lie algebroid structures. For complete conditions, an affine connection is needed.
}
Refer to Appendix \ref{connectionLA} for more details of 
Lie algebroid connections.

If we use the connections,
covariantized local coordinates denoted by attaching 
${}^\nabla$
in upper indices are defined by
\begin{align}
x^{\nabla i} &:= x^i,
\\
q^{\nabla i} &:= q^i,
\qquad p^{\nabla}_i := p_i,
\\
a^{\nabla a} &:= a^a,
\qquad
w^{\nabla}_a := w_a,
\\
z^{\nabla}_i &
:= z_i - \Gamma_{ji}^k p_k q^j - \omega_{ai}^b w_b a^a.
\end{align}
As in the above equations, only $z^{\nabla}_i$ is nontrivially redefined.
In fact, one can check that $z^{\nabla}_i $ transforms covariantly under 
local coordinate transformations, 
\eqref{coordinatetransf1}--\eqref{coordinatetransf6}.
For simplicity, only $z^{\nabla}_i$ is denoted by attaching ${}^\nabla$ 
in later sections.

The Liouville $1$-form $\vartheta$ for the graded symplectic form 
such that $\gomega = - \delta \vartheta$ is written by covariant coordinates 
as
\begin{align}
\vartheta &= 
(-1)^n z_i \delta x^i 
+ (-1)^{n-1} p_i \delta q^i 
+ (-1)^{n-1} w_a \delta a^a 
\nonumber \\
& = (-1)^n z_i^{\nabla} \delta x^i 
+ (-1)^{n-1} p_i D q^i 
+ (-1)^{n-1} w_a D a^a 
\label{Liouville3}
\end{align}
where $D q^i = \delta q^i - \Gamma_{kj}^i q^k \delta x^j$
and $D a^a = \delta a^a - \omega^a_{bi} a^b \delta x^i$
are covariant derivatives.

Nonzero Poisson brackets for covariant local coordinates are
\begin{align}
\sbv{x^i}{z_i^{\nabla}} &= - \sbv{z_j^{\nabla}}{x^i} = \delta^i_j,
\\
\sbv{q^i}{p_j} &= (-1)^n \sbv{p_j}{q^i} = \delta^i_j.
\\
\sbv{a^a}{w_b} &= (-1)^n \sbv{w_b}{a^a} = \delta^a_b,
\\
\sbv{z_i^{\nabla}}{q^j} &= - \sbv{q^j}{z_i^{\nabla}} = \Gamma_{ki}^j q^k,
\\
\sbv{z_i^{\nabla}}{p_j} &= - \sbv{p_j}{z_i^{\nabla}}
= - \Gamma_{ji}^k p_k,
\\
\sbv{z_i^{\nabla}}{a^a} &= -
\sbv{a^a} {z_i^{\nabla}} = \omega_{bi}^a a^b,
\\
\sbv{z_i^{\nabla}}{w_a} &= - \sbv{w_a}{z_i^{\nabla}}
= - \omega_{ai}^b w_b,
\\
\sbv{z_i^{\nabla}}{z_j^{\nabla}} &= R_{ijk}^l p_l q^k
+ F_{ija}^b w_b a^a,
\end{align}
where $R_{ijk}^l$ is the curvature for the affine connection
$\Gamma_{ij}^k$, 
and $F_{ija}^b$ is the curvature for the vector bundle connection
$\omega_{ai}^b$
given by 
\begin{align}
R_{ijk}^l & = \partial_i \Gamma_{kj}^l - \partial_j \Gamma_{ki}^l 
- \Gamma_{ki}^m \Gamma_{mj}^l + \Gamma_{kj}^m \Gamma_{mi}^l,
\\
F_{ija}^b &= \partial_i \omega_{aj}^b - \partial_j \omega_{ai}^b
- \omega_{ai}^c \omega_{cj}^b + \omega_{aj}^c \omega_{ci}^b.
\end{align}

\subsection{Homological functions}\label{}

Covariantized homological functions for 
non covariant functions
\eqref{homological11} and \eqref{homological12} 
are obtained by replacing $z_i$ to $z_i^{\nabla}$,
\begin{align}
\Thetas^{\nabla} &:= 
z^{\nabla}_i q^i
+ \frac{1}{(n+1)!} H_{i_1 \ldots i_{n+1}}(x) q^{i_1}
\ldots q^{i_{n+1}}
= \Thetas - \Gamma_{ji}^k(x) p_k q^j q^i
- \omega_{ai}^b(x) w_b a^a q^i,
\label{thetas01} 
\\
\Thetaa^{\nabla} &:= 
\rho^i_a(x) z^{\nabla}_i a^a - \frac{1}{2} T_{ab}^c(x) a^a a^b w_c
= \Thetaa 
- \rho^i_a(x) \Gamma_{ji}^k(x) p_k q^j a^a.
\label{thetaa01}
\end{align}
They are globally well-defined since all local coordinates and coefficient
quantities are covariant and tensors,
where $T_{ab}^c$ is the $A$-torsion on $A$.

The Poisson bracket $\sbv{-}{-}$ is induced from 
the symplectic form \eqref{gsymplectic3}
on $T^*[n](T[1]M \oplus A[1])$.
Since $\Thetas^{\nabla} \neq \Thetas$ and
$\Thetaa^{\nabla} \neq \Thetaa$,
homological conditions do not hold, i.e.,
$\sbv{\Thetas^{\nabla}}{\Thetas^{\nabla}} \neq 0$ and
$\sbv{\Thetaa^{\nabla}}{\Thetaa^{\nabla}} \neq 0$.
As a result, the total function
$\Thetag^{\nabla} := \Thetas^{\nabla} + \Thetaa^{\nabla}$
does not satisfy the homological condition, i.e.,
$\sbv{\Thetas^{\nabla}}{\Thetas^{\nabla}} \neq 0$.

In fact, the graded Poisson bracket
$\sbv{\Thetag^{\nabla}}{\Thetag^{\nabla}}$ is concretely calculated 
as follows,
\begin{align}
\sbv{\Thetas^{\nabla}}{\Thetas^{\nabla}} 
& = R_{ijk}^l p_l q^i q^j q^k + F_{ija}^b w_b a^a q^i q^j
+ 2 (-1)^n \Gamma_{kj}^i z_i^{\nabla} q^j q^k
\nonumber \\ & 
+ 2 \frac{(-1)^{n-1}}{(n+2)!} \partial_{[i_0} H_{i_1 \ldots i_{n+1}]}(x) q^{i_0} \ldots q^{i_{n+1}},
\label{cme01} \\
\sbv{\Thetaa^{\nabla}}{\Thetaa^{\nabla}} 
&= \rho^i_a \rho^j_b R_{ijk}^l p_l q^k a^a a^b,
\label{cme02} \\
\sbv{\Thetas^{\nabla}}{\Thetaa^{\nabla}} 
&=
- \nabla_i \rho_a^j z_j^{\nabla} q^i a^a
+ (-1)^{n-1} \rho^i_a (R_{ijk}^l q^j p_l q^k + F_{ijb}^c q^j w_c a^b) a^a
\nonumber \\ & 
- \frac{1}{2} (S_{iab}^c - 2 \rho^j_{[a} F_{ij|b]}^c) q^i a^a a^b w_c
 + \frac{1}{(n-1)!} \rho^j_a \nabla_{j} H_{i_1 \ldots i_{n+1}]}(x) q^{i_1} \ldots q^{i_{n+1}} a^a.
\label{cme03}
\end{align}
where local coordinate expressions of the curvature \eqref{curvatureona}, 
the $A$-torsion \eqref{atorsionona}, 
the basic curvature \eqref{bcurvatureona}
and the identity Eq.~\eqref{LAidentity1} have been used.
The total homological function for the gauged target space
is $\Thetag^{\nabla} = \Thetas^{\nabla} + \Thetaa$.
Therefore Eqs.~\eqref{cme01}--\eqref{cme03} leads us
that $\sbv{\Thetag^{\nabla}}{\Thetag^{\nabla}}=0$ is satisfied
if and only if the following conditions hold,
\begin{align}
& R_{ijk}^l = F_{ija}^b = S_{iab}^c =0,
\label{homological01}
\\
& \Gamma_{ij}^k - \Gamma_{ji}^k =0,
\label{homological03}
\\
& \nabla_i \rho^j_a =0,
\label{homological05}
\\
& \partial_{[i_0} H_{i_1\ldots i_{n+1}]} = 0.
\label{homological06}
\\
& \rho^j_{a} \nabla_{j} H_{i_1\ldots i_{n+1}} = 0.
\label{homological07}
\end{align}
By combining Eqs.~\eqref{homological05} and \eqref{homological07},
we obtain the equation ${}^A \rd^{\nabla} H = 0$.
The we summarize the result as follows.
\begin{theorem}\label{GHomologicalth}
A function $\Thetag^{\nabla}$ is homological $\sbv{\Thetag^{\nabla}}{\Thetag^{\nabla}}=0$ 
if and only if
$R = F = \baS =0$,
the anchor map is horizontal $\nabla \rho = 0$,
and $H$ satisfies $\rd H = {}^A \rd^{\nabla} H =0$.
\end{theorem}
In fact, conditions in Theorem \ref{GHomologicalth} are rather strong ones.
Because the Lie algebroid is almost fixed by the following proposition.\cite{Blaom}
\begin{proposition}\label{actLA}
A Lie algebroid $E$ equipped with a connection $\nabla$ such that 
$F = \baS = 0$, it is 
locally isomorphic to an action Lie algebroid $M \times \mathfrak{g}$
with a Lie algebra $\mathfrak{g}$.
\end{proposition}
From the above data, an AKSZ action $\SactG$ is constructed.
Let $(\ba^a, \bw_a)$ be superfields of degree $(1, n-1)$ on the fiber 
of $\Map(T[1]\Sigma, T^*[n]A[1])$
corresponding coordinates $(a^a, w_a)$.
The kinetic term is induced from the Liouville $1$-form 
\eqref{Liouville3} as
\begin{align}
& \int_{T[1]\Sigma} \rd \lambda_{n+1}
\left[(-1)^{n+1} \bz_i \rd \bX^i 
+ (-1)^n \bp_i \rd \bq^i
+ (-1)^n  \bw_a \rd \ba^a \right]
\nonumber \\ &=
\int_{T[1]\Sigma} \rd \lambda_{n+1}
\left[(-1)^{n+1}  \bz_i^{\nabla} \rd \bX^i 
+ (-1)^n \bp_i D \bq^i
+ (-1)^n \bw_a D \ba^a \right],
\end{align}
where $D \bq^i = \rd \bq^i - \Gamma^i_{kj} \bq^k \rd \bX^j$.
and $D \ba^a = \rd \ba^a - \omega^a_{bi} \ba^b \rd \bX^i$.
Remaining terms are induced from homological functions
\eqref{thetas01}  and \eqref{thetaa01} by transgression maps.
As a covariant modification of Eq.~\eqref{HCSMAKSZaciton},
the covariantized higher Courant sigma model is given by
\begin{align}
S_S^{\nabla} &= \int_{T[1]\Sigma} \rd \lambda_{n+1}
\left[
(-1)^{n+1}  \bz_i^{\nabla} \rd \bX^i 
+ (-1)^n \bp_i D \bq^i
+ \bz_i^{\nabla} \bq^i
\right.
\nonumber \\ &
\left.
+ \frac{1}{(n+1)!} H_{i_1 \ldots i_{n+1}}(\bX) \bq^{i_1} \cdots \bq^{i_{n+1}} 
\right],
\label{actionGHCSM01}
\end{align}
We add the following gauging terms induced from 
the Liouville $1$-form and $\Thetaa$,
\begin{align}
S_A^{\nabla} &= \int_{T[1]\Sigma} \rd \lambda_{n+1}
\left[(-1)^n \bw_a D \ba^a
+ \rho^i_a(\bX) \bz_i^{\nabla} \ba^a 
- \frac{1}{2} T_{ab}^c(\bX) \ba^a \ba^b \bw_c \right].
\label{actionGHCSM02}
\end{align}
The resulting AKSZ action of a gauged higher Courant Sigma model 
(GHCSM) with Lie algebroid gauging is
the sum of Eq.~\eqref{actionGHCSM01}
and Eq.~\eqref{actionGHCSM02},
\begin{align}
\SactG &= S_S^{\nabla} + S_A^{\nabla}
\nonumber \\ 
&=\int_{T[1]\Sigma} \rd \lambda_{n+1}
\left[(-1)^{n+1}  \bz_i^{\nabla} \rd \bX^i 
+ (-1)^n \bp_i D \bq^i
+ (-1)^n \bw_a D \ba^a
+ \bz_i^{\nabla} \bq^i
\right.
\nonumber \\
& 
\left.
+ \frac{1}{(n+1)!} H_{i_1 \ldots i_{n+1}}(\bX) \bq^{i_1} \cdots \bq^{i_{n+1}} 
+ \rho^i_a(\bX) \bz_i^{\nabla} \ba^a 
- \frac{1}{2} T_{ab}^c(\bX) \ba^a \ba^b \bw_c \right].
\label{actionSCSMLA}
\end{align}
BY applying Theorem \ref{GHomologicalth} to the sigma model 
\eqref{actionSCSMLA}, we obtain the following,
\begin{corollary}\label{GHomologicalth2}
The BV action functional $\SactG$ satisfies the classical master equation 
$\sbv{\SactG}{\SactG}=0$ 
if and only if
$R = F = \baS =0$,
the anchor map is horizontal $\nabla \rho = 0$,
and $H$ satisfies $\rd H = {}^A \rd^{\nabla} H =0$.
\end{corollary}

\begin{example}
If gauging symmetry is a Lie group, the infinitesimal symmetry is 
a Lie algebra. Let $\mathfrak{g}$ be Lie algebra of the symmetry.
Then, $A = M \times \mathfrak{g}$ is an action Lie algebroid
and the affine connection and the vector bundle connection are
trivial $\nabla = \rd$.
Then $C_{ab}^c = - T_{ab}^c $ is globally constant, i.e., 
the structure constant of $\mathfrak{g}$, 
and a vector field $\rho_a := \rho^i_a(x)\partial_i$ 
is the infinitesimal action of $\mathfrak{g}$ as a differential operator.
$\SactG$ reduces the action of a simpler nonabelian BF type gauge theory.
All the conditions in Theorem \ref{GHomologicalth} are satisfied, thus, 
$\sbv{\Thetag^{\nabla}}{\Thetag^{\nabla}}=0$ holds and 
the sigma model is consistent without any extra geometric condition.
\end{example}

\section{GHCSMs with fluxes and boundaries}\label{sec:HCSMflux}
\noindent
In this section, we consider extensions of 
the gauged higher Courant sigma model (GHCSM) in Section \ref{sec:GHCSM}
by introducing flux terms and boundary terms.
On target bundles, New geometric structures appear. They are generalizations 
of momentum maps and Hamiltonian $G$-spaces.

\subsection{Introduction of fluxes}\label{sec:SCSMLAflux}
The homological function
$\Thetag^{\nabla} = \Thetas^{\nabla} + \Thetaa^{\nabla}$, 
with Eq.~\eqref{thetas01} and \eqref{thetaa01} is not the most general one.
More terms of degree $n+1$ can be added to $\Thetag^{\nabla}$.

In this section, we add the following terms called flux terms,
\begin{align}
\Theta_F &=
\sum_{s=0}^{n} \frac{1}{s!(n+1-s)!} 
H^{(s)}_{i_1 \ldots i_s a_{s+1} \ldots a_{n+1}}(x) q^{i_1} \cdots q^{i_s} 
a^{a_{s+1}} \cdots a^{a_{n+1}},
\label{thetaf}
\end{align}
to $\Thetag^{\nabla}$.
Here $H^{(s)} = 
\frac{1}{s!(n+1-s)!} 
H^{(s)}_{i_1 \ldots i_s a_{s+1} \ldots a_{n+1}}(x) \rd x^{i_1} \wedge 
\cdots \wedge \rd x^{i_s} \otimes 
e^{a_{s+1}} \wedge \cdots \wedge e^{a_{n+1}} 
\in \Omega^s(M, \wedge^{n+1-s} A^*)$ 
is an $s$-form taking a value in $\wedge^{n+1-s} A^*$.
Here $e^a$ is a basis of $A^*$.
More general degree $n+1$ terms induced from functions $H(x, q, p, a, w)$ 
can be assumed. However in this paper, we add the term in Eq.~\eqref{thetaf}
since this includes interesting structures.

The total homological function is 
$\Thetagf := \Thetas^{\nabla} + \Thetaa^{\nabla} + \Theta_F$.
Deforming the homological function, consistent target geometry is deformed.

New Poisson brackets including $\Thetaf$ are as follows,
\footnote{Notation: $\partial_{[i} \alpha_{j]}
= \partial_{i} \alpha_{j}- \partial_{j} \alpha_{i}$,
and 
$\partial_{[i} T_{jk]}
= \partial_{i} T_{jk} + \partial_{j} T_{ki} + \partial_{k} T_{ij}$.
for an antisymmetric tensor $T_{ij}$, etc. }
\begin{align}
& \sbv{\Thetas^{\nabla}}{\Thetaf} 
\nonumber \\
& = 
- \sum_{s=0}^{n} \frac{1}{(s+1)!(n+1-s)!} 
\nabla_{[i_0} H^{(s)}_{i_1 \ldots i_s] 
a_{s+1} \ldots a_{n+1}}(x) q^{i_0} q^{i_1} \cdots q^{i_s} 
a^{a_{s+1}} \cdots a^{a_{n+1}},
\label{cme04}
\\
& \sbv{\Thetaa^{\nabla}}{\Thetaf} 
= 
- \sum_{s=0}^{n} \frac{1}{(s+1)!(n+2-s)!} 
\left(
\rho^j_{[a_{s+1}|} \nabla_{j} H^{(s)}_{i_1 \ldots i_s
|a_{s+2} \ldots a_{n+2}}(x) 
\right.
\nonumber \\ 
& 
\left.
+ (-1)^{n-s}
T^b_{[a_{s+1}a_{s+2}|} H^{(s)}_{i_1 \ldots i_s] 
b |a_{s+3} \ldots a_{n+2}]}(x)
\right) q^{i_1} \cdots q^{i_s} 
a^{a_{s+1}} \cdots a^{a_{n+2}},
\label{cme05} 
\\
& \sbv{\Thetaf}{\Thetaf}  = 0.
\label{cme06}
\end{align}
If $\sbv{\Thetagf}{\Thetagf} =0$ is imposed for 
the deformed function,
$\Thetagf = \Thetas^{\nabla} + \Thetaa^{\nabla} + \Theta_F$,
fluxes $H^{(s)}$ must satisfy the following conditions 
obtained from Eqs.~\eqref{cme04}-\eqref{cme06}.
Here the right hand of Eq.~\eqref{cme04} is the exterior covariant derivative
$\nabla H^{(s)}$, and 
the right hand side of Eq.~\eqref{cme05} is
the Lie algebroid differential ${}^A \rd H^{(s)}$.
Under $\nabla \rho = 0$, ${}^A \rd H^{(s)}$ is equivalent to 
the $A$-exterior covariant derivative
${}^A \rd = {}^A \rd^{\nabla}$ for the basic $A$-connection.
Combining Eqs.~\eqref{cme04}-\eqref{cme06} 
with Eqs.~\eqref{cme01}-\eqref{cme03},
we obtain the following result.
\begin{theorem}\label{SCSMLAflux}
$\Theta^{\nabla} = \Thetas + \Thetaa + \Thetaf$
is homological if and only if the following conditions are satisfied,
\begin{align}
& R = F = \baS =0,  \quad \nabla \rho = 0,
\label{fluxeq01}
\\
& \rd H = 0,
\label{fluxeq02}
\\
& {}^A \rd^{\nabla} H^{(s+1)} - \nabla H^{(s)} = 0,
\qquad \mbox{for} \ s = 0, \ldots n 
\label{fluxeq03}
\\
& {}^A \rd^{\nabla} H^{(0)} =0.
\label{fluxeq04}
\end{align}
\end{theorem}
\begin{example}\label{fluxsol}
From the condition $\nabla \rho = 0$ in Theorem \ref{GHomologicalth}
and the identity of the Lie algebroid Eq.~\eqref{LAidentity01},
$T = 0$ or $\rho=0$.
Choose the solution $T = 0$, i.e., the vanishing $A$-torsion.
Then, one solution of Eqs.~\eqref{fluxeq03}--\eqref{fluxeq04}
is $H^{(s)} = \iota_{\rho}^{n+1-s} H$, where
$H = H^{(n+1)}$ is assumed to satisfy Eq.~\eqref{fluxeq02}.
For a given closed $n+1$-form $H \in \Omega^{n+1}(M)$,
where $\iota_{\rho}(e_{n+1-s}) \ldots 
\iota_{\rho}(e_1) H
:= H(\rho(e_1), \ldots, \rho(e_{n+1-s}))$ 
for $e_1, \ldots, e_{n+1-s} \in \Gamma(A)$, 
etc.
\end{example}
The AKSZ action corresponding to the 
function $\Thetaf$ in Eq.~\eqref{thetaf} 
the following term,
\begin{align}
S_F
&= \sum_{s=0}^{n} 
\frac{1}{s!(n+1-s)!} 
\int_{T[1]\Sigma} \rd \lambda_{n+1}
H^{(s)}_{i_1 \ldots i_s a_{s+1} \ldots a_{n+1}}(\bX) \bq^{i_1} 
\cdots \bq^{i_s} 
\ba^{a_{s+1}} \cdots \ba^{a_{n+1}}.
\label{SCSMLAfluxaction}
\end{align}
Therefore the total action is 
\begin{align}
\SactGF &= S_S^{\nabla} + S_A^{\nabla} + S_F
\nonumber \\ 
&=\int_{T[1]\Sigma} \rd \lambda_{n+1}
\left[(-1)^{n+1}  \bz_i^{\nabla} \rd \bX^i 
+ (-1)^n \bp_i D \bq^i
+ (-1)^n \bw_a D \ba^a
+ \bz_i^{\nabla} \bq^i
\right.
\nonumber \\
& 
\left.
+ \frac{1}{(n+1)!} H_{i_1 \ldots i_{n+1}}(\bX) \bq^{i_1} \cdots \bq^{i_{n+1}} 
+ \rho^i_a(\bX) \bz_i^{\nabla} \ba^a 
- \frac{1}{2} T_{ab}^c(\bX) \ba^a \ba^b \bw_c 
\right.
\nonumber \\
& 
\left.
+ \sum_{s=0}^{n} 
\frac{1}{s!(n+1-s)!} 
H^{(s)}_{i_1 \ldots i_s a_{s+1} \ldots a_{n+1}}(\bX) \bq^{i_1} 
\cdots \bq^{i_s} 
\ba^{a_{s+1}} \cdots \ba^{a_{n+1}}
\right].
\label{actionSCSMLAflux}
\end{align}
From Theorem \ref{SCSMLAflux}, the following corollary is obtained.
\begin{corollary}
If and only if the conditions in Theorem \ref{SCSMLAflux} are satisfied,
the action functional \eqref{actionSCSMLAflux} satisfies the classical master 
equation $\sbv{\SactGF}{\SactGF}=0$.
\end{corollary}

\subsection{GHCSMs with boundaries}\label{sec:SCSMLAboundary}
\noindent
In this section, we consider a generalization to the GHCSM with boundaries.
Suppose that the source manifold $\Sigma$ 
has boundaries $\partial \Sigma \neq \emptyset$.
Then there exists extra freedom to add boundary terms to the action functional.
We discuss consistency of boundary conditions 
with homological conditions.

Take the action functional of the gauged higher Courant
sigma model with fluxes and Lie algebroid gauging \eqref{actionSCSMLAflux}
as a bulk theory on $\Sigma$. 
Since boundaries $\partial \Sigma$ is in $n$-dimensions,
boundary terms are integrations of the pullback of degree $n$ functions 
on $\calM$. Here, we consider following boundary terms,
\begin{align}
S_b &=
\sum_{s=0}^{n} \frac{1}{s!(n-s)!} 
\int_{T[1]\partial \Sigma} \rd \lambda_n
\mu^{(s)}_{i_1 \ldots i_s a_{s+1} \ldots a_{n}}(\bX) 
\bq^{i_1} \cdots \bq^{i_s} \ba^{a_{s+1}} \cdots \ba^{a_{s+n}},
\label{boundaryterms}
\end{align}
where 
$\mu^{(s)} := \frac{1}{s!(n-s)!}
\mu^{(s)}_{i_1 \ldots i_s a_{s+1} \ldots a_{n}}(x) 
\rd x^{i_1} \wedge \cdots \wedge \rd x^{i_s} 
\otimes e^{a_{s+1}} \wedge \cdots \wedge e^{a_{s+n}}
$ 
is an $s$-form taking a value in $\wedge^{n-s}A^*$,
$\mu^{(s)} \in \Omega^{s}(M, \wedge^{n-s}A^*)$.

Note that we can consider more general terms including $p_i, w_a$.
The choice \eqref{boundaryterms} is interesting since 
these includes interesting geometric structures,
and regarded as a generalization of boundary conditions
in gauged Poisson sigma models \cite{Ikeda:2023pdr}
and Courant sigma models with boundary in \cite{Ikeda:2012pv}.
In the case of the CSM, boundary geometry is described by 
Dirac structures such as the twisted Poisson structure 
\cite{Klimcik:2001vg, Park2000au}.
For gauged Poisson sigma models,
boundary terms are described by momentum maps or momentum sections on 
Poisson manifolds \cite{Ikeda:2023pdr}.

Let $\SactGF = S_S^{\nabla} + S_A^{\nabla} + S_F$ be the bulk 
action functional with 
$S_S^{\nabla} + S_A^{\nabla}$ in Eq.~\eqref{actionSCSMLAflux}
and $S_F$ in Eq.~\eqref{SCSMLAfluxaction},
and $\SactT = \SactGF + S_b$ be the total BV action functional
including boundary terms.
If $\Sigma$ has boundaries, $\sbv{\SactGF}{\SactGF}$ 
for the bulk action $\SactGF$ gives rise to boundary terms.
Especially, the Poisson bracket does not vanish. 
Concrete calculations give the following boundary integrations,
\begin{align}
& \sbv{\SactGF}{\SactGF} 
\nonumber \\ 
&= 2 \int_{T[1]\Sigma} \rd \lambda_{n+1} \
\rd 
\left[(-1)^{n+1}  \bz_i^{\nabla} \rd \bX^i 
+ (-1)^n \bp_i D \bq^i
+ (-1)^n \bw_a D \ba^a
+ \bz_i^{\nabla} \bq^i
\right.
\nonumber \\
& 
\left.
+ \frac{1}{(n+1)!} H_{i_1 \ldots i_{n+1}}(\bX) \bq^{i_1} \cdots \bq^{i_{n+1}} 
+ \rho^i_a(\bX) \bz_i^{\nabla} \ba^a 
- \frac{1}{2} T_{ab}^c(\bX) \ba^a \ba^b \bw_c 
\right.
\nonumber \\
& 
\left.
+ \sum_{s=0}^{n} 
\frac{1}{s!(n+1-s)!} 
H^{(s)}_{i_1 \ldots i_s a_{s+1} \ldots a_{n+1}}(\bX) \bq^{i_1} 
\cdots \bq^{i_s} 
\ba^{a_{s+1}} \cdots \ba^{a_{n+1}}
\right]
\nonumber \\ 
&= 2 \int_{T[1]\partial \Sigma} \rd \lambda_n
\left[(-1)^{n+1}  \bz_i^{\nabla} \rd \bX^i 
+ (-1)^n \bp_i D \bq^i
+ (-1)^n \bw_a D \ba^a
+ \bz_i^{\nabla} \bq^i
\right.
\nonumber \\
& 
\left.
+ \frac{1}{(n+1)!} H_{i_1 \ldots i_{n+1}}(\bX) \bq^{i_1} \cdots \bq^{i_{n+1}} 
+ \rho^i_a(\bX) \bz_i^{\nabla} \ba^a 
- \frac{1}{2} T_{ab}^c(\bX) \ba^a \ba^b \bw_c 
\right.
\nonumber \\
& 
\left.
+ \sum_{s=0}^{n} 
\frac{1}{s!(n+1-s)!} 
H^{(s)}_{i_1 \ldots i_s a_{s+1} \ldots a_{n+1}}(\bX) \bq^{i_1} 
\cdots \bq^{i_s} 
\ba^{a_{s+1}} \cdots \ba^{a_{n+1}}
\right].
\label{SGFSGF}
\end{align}
Here total derivative terms become integrations on boundaries using the Stokes' theorem.

In order to compute $\sbv{\SactT}{\SactT}$,
we use formulas $\sbv{\SactGF}{\Phi}$ for every fundamental superfield $\Phi$. 
They are
\begin{align}
\delta \bX^i &= \sbv{\SactGF}{\bX^i} = \rd \bX^i 
+ (-1)^{n} \bq^i + (-1)^{n} \rho^i_a(\bX) \ba^a,
\label{bvofx}
 \\
\delta \bq^i &= \sbv{\SactGF}{\bq^i} = \rd \bq^i,
\label{bvofq} \\
\delta \ba^a &= \sbv{\SactGF}{\ba^a} = \rd \ba^a + 
(-1)^n \omega^a_{bi}(\bX) \ba^b \bq^i 
+ \frac{(-1)^{n-1}}{2} C^a_{bc}(\bX) \ba^b \ba^c.
\label{bvofa}
\end{align}
Using Eq.~\eqref{bvofx}--\eqref{bvofa}, the Poisson bracket
$\sbv{\SactGF}{S_b}$ is computed as
\begin{align}
\delta S_b &= \sbv{\SactGF}{S_b}
\nonumber \\ 
&= - \int_{T[1]\partial \Sigma} \rd \lambda_n
\left[\frac{1}{(n+1)!} (\rd \mu^{(n)})_{i_0 \ldots i_n}
\bq^{i_0} \cdots \bq^{i_n}
\right.
\nonumber \\ & \quad
\left.
- \sum_{s=0}^{n-1} \frac{1}{(s+1)!(n-s)!} 
\left(({}^A \rd \mu^{(s+1)})_{i_1 \ldots i_{s+1}a_{s+2} \ldots a_{n+1}}
+ (\nabla \mu^{(s)})_{i_1 \ldots i_{s+1}a_{s+2} \ldots a_{n+1}}
\right) 
\right.
\nonumber \\ & \quad
\left.
\times 
\bq^{i_1} \cdots \bq^{i_{s+1}} \ba^{a_{s+2}} \cdots \ba^{a_{n+1}}
+ \frac{1}{(n+1)!} ({}^A \rd \mu^{(0)})_{a_0 \ldots a_n} 
\ba^{a_0} \cdots \ba^{a_n}
\right],
\label{SGFSb}
\\
& \sbv{S_b}{S_b} = 0.
\label{SbSb}
\end{align}
We impose the homological condition $\sbv{\SactT}{\SactT}=0$ for consistency.
From the general theory in Section \ref{sec:GAKSZb}, 
the image of $T[1]\partial \Sigma$ is in
a coisotropic submanifold of the target QP-manifold
$\calN = T^*[n](T[1]M \oplus A[1])$.
In this model, we take simple boundary conditions,
the Lagrangian submanifold $\calL$ of $\calM$ satisfying 
$\bz_i = 0$,
$\bp_i =0$ and $\bw_a =0$ on $T[1]\partial\Sigma$.
These boundary conditions give interesting 
geometric structures on the target space 
on the corresponding non graded 
ordinary vector bundle $TM \oplus \wedge^{n-1} T^*M \oplus A$.

Moreover the condition \eqref{boundarymaster} is imposed.
The equation \eqref{boundarymaster} gives identities between $H^{(s)}$
and $\mu^{(s)}$. Substituting concrete calculations,
\eqref{SGFSGF}, \eqref{SGFSb} and \eqref{SbSb},
we obtain the following theorem.
%
%
%
%
\begin{theorem}
Take boundary conditions $\bz_i = 0$,$\bp_i =0$ and $\bw_a =0$ 
on $T[1]\partial\Sigma$. Then,
$\SactT$ is homological, i.e., 
satisfies the classical master equation if and only if
in addition to equations in Theorem \ref{SCSMLAflux},
the following conditions are satisfied,
\begin{align}
& \rd \mu^{(n)} - H = 0,
\label{hmm01}
\\
& {}^A \rd^{\nabla} \mu^{(s+1)} + \nabla \mu^{(s)} - H^{(s+1)} = 0,
\qquad \mbox{for} \ s = 0, \ldots n-1,
\label{hmm02}
\\
& {}^A \rd^{\nabla} \mu^{(0)} - H^{(0)} = 0.
\label{hmm03}
\end{align}
\end{theorem}
There is a geometric interpretation of Eqs.~\eqref{hmm01}--\eqref{hmm03}.
Elements $\mu^{(s)}$ satisfying \eqref{hmm01}--\eqref{hmm03}
are generalizations of homotopy momentum sections on 
a (pre-)$n$-plectic manifold. \cite{Hirota:2021isx}
In fact, we have the following example.
\begin{example}
Assume Eq.~\eqref{fluxeq01} and the solution in Example \ref{fluxsol},
$H^{(s)} = (\iota_{\rho})^{n+1-s} H$.
Then, $\mu^{(s)}$'s satisfying Eqs.~\eqref{hmm01}--\eqref{hmm03} 
are homotopy momentum sections. \cite{Hirota:2021isx}
Moreover if the manifold is a (pre-)$n$-plectic manifold
with a Lie group action and 
a Lie algebroid $A$ is an action Lie algebroid 
$A = M \times \mathfrak{g}$ induced from the Lie group action,
conditions in Theorem \ref{SCSMLAflux} and $T=0$ are satisfied.
Then, if the action $\rho$ is equivariant,
the $A$-differential ${}^A \rd$ reduces to 
the Chevalley-Eilenberg differential of the Lie algebra $d_{CE}$
and $\mu^{(s)}$'s are homotopy moment maps \cite{Callies:2013jbu}.
In fact, Eqs.~\eqref{hmm01}--\eqref{hmm03} reduce to the definition of 
homotopy moment maps. \cite{Hirota:2021isx}
\end{example}
Eqs.~\eqref{hmm01}--\eqref{hmm03} are regarded as a generalization of 
homotopy momentum maps to
higher Courant algebroids with Lie algebroid actions.

\subsection*{Acknowledgments}
\noindent
The author would like to thank Simon-Raphael Fischer
for useful comments and discussions.

The author is grateful to the Institut {\'E}lie Cartan de Lorraine for supports, where part of this work was carried, for their hospitality.

This work was supported by JSPS Grants-in-Aid for Scientific Research Number 22K03323.

\appendix

\section{Lie algebroids}\label{app:Liealgebroid}
In Appendix, we summarize definition and formulas of Lie algebroids, 
Courant algebroids and graded manifolds.

\subsection{Definitions and examples}

\begin{definition}
Let $A$ be a vector bundle over a smooth manifold $M$.
A Lie algebroid $(A, [-,-], \rhoa)$ is a vector bundle $A$ with
a bundle map $\rho: A \rightarrow TM$ called the anchor map, 
and a Lie bracket
$[-,-]: \Gamma(A) \times \Gamma(A) \rightarrow \Gamma(A)$
satisfying the Leibniz rule,
\begin{eqnarray}
[e_1, fe_2] &=& f [e_1, e_2] + \rho(e_1) f \cdot e_2,
\end{eqnarray}
where $e_1, e_2 \in \Gamma(A)$ and $f \in C^{\infty}(M)$.
\end{definition}
A Lie algebroid generalizes both a Lie algebra and the space of vector fields on a smooth manifold.
\begin{example}[Lie algebras]
Let a manifold $M$ be one point $M = \{pt \}$. 
Then a Lie algebroid is a Lie algebra $\mathfrak{g}$.
\end{example}
\begin{example}[Tangent Lie algebroids]\label{tangentLA}
If a vector bundle $A$ is a tangent bundle $TM$ and $\rhoa = \mathrm{id}$, 
then a bracket $[-,-]$ is a normal Lie bracket 
on the space of vector fields $\mathfrak{X}(M)$
and $(TM, [-,-], \mathrm{id})$ is a Lie algebroid.
It is called a \textit{tangent Lie algebroid}.
\end{example}

\begin{example}[Action Lie algebroids]\label{actionLA}
Assume a smooth action of a Lie group $G$ to a smooth manifold $M$, 
{$M \times G \rightarrow M$.}
The differential map of this action induces an infinitesimal action of the Lie algebra $\mathfrak{g}$ of $G$ on $M$.
Since $\mathfrak{g}$ acts as a differential operator on $M$,
the differential map
is a bundle map $\rho: M \times \mathfrak{g} \rightarrow TM$.
Consistency of a Lie bracket requires that $\rho$ is 
a Lie algebra morphism such that
\begin{eqnarray}
~[\rho(e_1), \rho(e_2)] &=& \rho([e_1, e_2]),
\label{almostLA}
\end{eqnarray}
where the bracket on the left-hand side 
is the standard Lie bracket of vector fields.  
These data define a Lie algebroid $(A= M \times \mathfrak{g}, [-,-], \rho)$.
known as an \textit{action Lie algebroid}.
\end{example}

For a comprehensive review of Lie algebroids and their properties, see, 
for example, \cite{Mackenzie}.

For a Lie algebroid $A$, sections of the exterior algebra of $A^*$ are called \textit{$A$-differential forms}.
On the spaces of $A$-differential forms, $\Gamma(\wedge^{\bullet} A^*)$,
a differential ${}^A \rd: \Gamma(\wedge^m A^*)
\rightarrow \Gamma(\wedge^{m+1} A^*)$ 
called the \textit{Lie algebroid differential}, 
or the \textit{$A$-differential}, is defined as follows. 
\begin{definition}
The $A$-differential ${}^A \rd: \Gamma(\wedge^m A^*)
\rightarrow \Gamma(\wedge^{m+1} A^*)$ is defined by
\begin{eqnarray}
{}^A \rd \mu(e_1, \ldots, e_{m+1}) 
&=& \sum_{i=1}^{m+1} (-1)^{i-1} \rho(e_i) \mu(e_1, \ldots, 
\check{e_i}, \ldots, e_{m+1})
\nonumber \\ && 
+ \sum_{1 \leq i < j \leq m+1} (-1)^{i+j} \mu([e_i, e_j], e_1, \ldots, \check{e_i}, \ldots, \check{e_j}, \ldots, e_{m+1}),
\label{LAdifferential}
\end{eqnarray}
where $\mu \in \Gamma(\wedge^m A^*)$ and $e_i \in \Gamma(A)$.
\end{definition}
The $A$-differential satisfies $({}^A \rd)^2=0$.
It generalizes both the de Rham differential on $T^*M$ and the Chevalley-Eilenberg differential on a Lie algebra.

\subsection{Connections on Lie algebroids}\label{connectionLA}
Two types of connections are defined on Lie algebroids.
We start by introducing an ordinary connection on a vector bundle $A$.
A connection is an $\bR$-linear map,
$\nabla:\Gamma(A)\rightarrow \Gamma(A \otimes T^*M)$,
satisfying the Leibniz rule,
\begin{eqnarray}
\nabla (f e) = f \nabla e + (\rd f) \otimes e,
\end{eqnarray}
for every $e \in \Gamma(A)$ and $f \in C^{\infty}(M)$.
The dual connection on $A^*$ is defined by the equation,
\begin{eqnarray}
\rd \inner{\mu}{e} = \inner{\nabla \mu}{e} + \inner{\mu}{\nabla e},
\end{eqnarray}
for all sections $\mu \in \Gamma(A^*)$ and $e \in \Gamma(A)$,
where $\inner{-}{-}$ denotes the pairing between $A$ and $A^*$.
For simplicity, we use the same notation $\nabla$ for the dual connection.

On a Lie algebroid, we define another derivation called an $A$-connection.
Let $E$ be another vector bundle over the same base manifold $M$.
An \textit{$A$-connection} on $E$ 
with respect to the Lie algebroid $A$ is an $\bR$-linear map,
${}^A \nabla: \Gamma(E) \rightarrow \Gamma(E \otimes A^*)$,
satisfying 
\begin{eqnarray}
{}^A \nabla_e (f s) = f {}^A \nabla_e s + (\rho(e) f) s,
\end{eqnarray}
for $e \in \Gamma(A)$, $s \in \Gamma(E)$ and $f \in C^{\infty}(M)$.
%
The ordinary connection on $TM$ is a special case of an 
$A$-connection for $A=TM$, the ordinary connection 
$\nabla$ corresponds to the $TM$-connection ${}^{TM} \nabla$
on $A=TM$.

If an ordinary vector bundle connection $\nabla$ on $A$ is given,
an $A$-connection called the \textit{basic $A$-connection} is defined.
The basic $A$-connection on $E=A$,
$\anablab: \Gamma(A) \rightarrow \Gamma(A \otimes A^*)$ 
is defined by
\begin{eqnarray}
\anablab_{e} e^{\prime} &:=& 
\nabla_{\rho(e)} {e^{\prime}} + [{e}, {e^{\prime}}]_A,
\label{basicAconnection2}
\end{eqnarray}
for $e, e^{\prime} \in \Gamma(A)$.
A basic $A$-connection on the tangent bundle $E=TM$,
${}^A \nabla: \Gamma(TM) \rightarrow \Gamma(TM \otimes A^*)$,
is defined by
\begin{eqnarray}
\anablab_{e} v &:=& \calL_{\rho(e)} v + \rho(\nabla_v e)
= [\rho(e), v] + \rho(\nabla_v e),
v
\label{stAconnection1}
\end{eqnarray}
where 
$e \in \Gamma(A)$ and $v \in \mathfrak{X}(M)$.
For a $1$-form $\alpha \in \Omega^1(M)$, the \textit{basic $A$-connection} 
is given by
\begin{eqnarray}
\anablab_{e} \alpha &:=& \calL_{\rho(e)} \alpha 
+ \inner{\rho(\nabla e)}{\alpha},
\label{Econoneform}
\end{eqnarray}
as the dual connection. 
For general discussions on connections on Lie algebroids, see
\cite{DufourZung, AbadCrainic, CrainicFernandes}.

Given a connection $\nabla$, the covariant derivative extends naturally  
to the derivation on the space of differential forms taking a value in
$\wedge^m A^*$,
$\Omega^k(M, \wedge^m A^*)$, which is called 
the \textit{exterior covariant derivative}.

Similarly, an $A$-connection ${}^A \nabla$ extends to a derivation
satisfying the Leibniz rule on the space 
$\Gamma(\wedge^m A^* \otimes \wedge^k E)$.
This extension is called \textit{the $A$-exterior covariant derivative}
${}^A \rd^{\nabla}: 
{}^A \nabla: \Gamma(\wedge^m A^* \otimes \wedge^k E)
\rightarrow \Gamma(\wedge^{m+1} A^* \otimes \wedge^k E)$
and is denoted by the same notation $\nabla$ and ${}^A\nabla$.
In this paper, we consider the $A$-exterior covariant derivatives 
on $E = T^*M$.
For $E = T^*M$, the concrete definition is as follows.
\begin{definition}
For $\Omega^k(M, \wedge^m A) = \Gamma(\wedge^m A^* \otimes \wedge^k T^*M)$,
the \textit{$A$-exterior covariant derivative}
${}^A \nabla: \Omega^k(M, \wedge^m A^*) \rightarrow \Omega^k(M, \wedge^{m+1} A^*)$ is defined by
\begin{eqnarray}
({}^A \rd^{\nabla} \alpha)(e_1, \ldots, e_{m+1})
&:=& \sum_{i=1}^{m+1} (-1)^{i-1} 
{}^A \nabla_{e_i}
(\alpha(e_1, \ldots, \check{e_i}, \ldots, e_{m+1}))
\nonumber \\ &+ & 
\sum_{1 \leq i < j \leq m+1} (-1)^{i+j} \alpha([e_i, e_j], e_1, \ldots, \check{e_i}, \ldots, \check{e_j}, \ldots, e_{m+1}),
\label{Ediffconnection}
\end{eqnarray}
for $\alpha \in \Omega^k(M, \wedge^m A^*)$ and $e_i \in \Gamma(A)$.
\end{definition}
For two connections $\nabla$ and ${}^A \nabla$, 
we introduce tensors:
the \textit{curvature}, $R \in \Omega^2(M, A \otimes A^*)$, 
the \textit{$A$-torsion}, $T \in \Gamma(A \otimes \wedge^2 A^*)$, 
and the \textit{basic curvature} \cite{Blaom}.
$\baS \in \Omega^1(M, \wedge^2 A^* \otimes A)$.
These are defined as follows:
\beqa
R(v, v^{\prime}) &:=& [\nabla_v, \nabla_{v^{\prime}}] - \nabla_{[v, v^{\prime}]}, 
\label{curv}
\\
T(e, e^{\prime}) &:=& \nabla_{\rho(e)} e^{\prime} - \nabla_{\rho(e^{\prime})} e
- [e, e^{\prime}],
\label{Etorsion}
\\
\baS(e, e^{\prime}) &:=& 
[e, \nabla e^{\prime}] - [e^{\prime}, \nabla e]
- \nabla[e, e^{\prime}] 
- \nabla_{\rho(\nabla e)} e^{\prime} + \nabla_{\rho(\nabla e^{\prime})} e
\nonumber \\ 
&=& (\nabla T + 2 \mathrm{Alt} \, \iota_\rho R)(e, e^{\prime}),
\label{bcurv}
\eeqa
for $v, v^{\prime} \in \mathfrak{X}(M)$ and $e, e^{\prime} \in \Gamma(A)$.
Note that $\iota_\rho R \in \Omega^1(M, A^* \otimes A^* \otimes A)$ since
$\rho \in \Gamma(A^* \otimes TM)$ and $\iota_\rho$ gives the contraction 
between $TM$ and $T^*M$. Notation $\mathrm{Alt}$ means skew symmetrization on $A^* \otimes A^*$, which gives an element in $\Omega^1(M, \wedge^2 A^* \otimes A)$

\subsection{Local coordinate expressions}
In this section, we list formulas in local coordinates in Lie algebroids .

For a Lie algebroid $A$ over $M$,
$x^i$ is a local coordinate on $M$, $e_a \in \Gamma(A)$ is a basis of sections of $A$ and $e^a \in \Gamma(A^*)$ is a dual basis of sections of $A^*$. 
$i,j$, etc.~are indices on $M$ and $a,b$, etc.~are indices on the fiber of $A$.
Local coordinate expressions of the anchor map and the Lie bracket are
$\rho(e_a) f = \rho^i_a(x) \partial_i f$ and
$[e_a, e_b ] = C_{ab}^c(x) e_c$, where $f \in C^{\infty}(M)$ and $\partial_i = \tfrac{\partial}{\partial x^i}$.
Then, identities of $\rho^i_a$ and $C_{ab}^c$ induced from the Lie algebroid conditions are
\beqa 
&& \rho_a^j \partial_j \rho_{b}^i - \rho_b^j \partial_j \rho_{a}^i = C_{ab}^c \rho_c^i,
\label{LAidentity1}
\\
&& C_{ad}^e C_{bc}^d + \rho_a^i \partial_i C_{bc}^e + \mbox{Cycl}(abc) = 0.
\label{LAidentity2}
\eeqa

Let $\omega = \omega^b_{ai} \rd x^i \otimes e^a \otimes e_b$ be 
a connection $1$-form for a connection on $A$, 
$\nabla:\Gamma(A) \rightarrow \Gamma(A \otimes T^*M)$.
For the basis vectors, the covariant derivatives are
${\nabla} e_a = - \omega_{ai}^b \rd x^i \otimes e_b$
and ${\nabla} e^a = \omega_{bi}^a \rd x^i \otimes e^b$.
Local coordinate expressions of 
covariant derivatives and the standard $A$-covariant derivatives 
on $TM$ and $A$, 
the basic $A$-connections are
\begin{eqnarray}
\nabla_i \alpha^a &=& \partial_i \alpha^a {+} \omega_{bi}^a \alpha^b,
\\
\nabla_i \beta_a &=& \partial_i \beta_a {-} \omega_{ai}^b \beta_b,
\\
\anablab_a v^i
&=& \rho_a^j \partial_j v^i - \partial_j \rho^i_a v^j
+ \rho^i_b \omega^b_{aj} v^j
= \rho_a^j \partial_j v^i - \nabla_j \rho^i_a v^j,
\\
\anablab_a \alpha_i
&=&  \rho_a^j \partial_j \alpha_i + \partial_i \rho^j_a \alpha_j
{-} \rho^j_b \omega^b_{ai} \alpha_j = 
\rho_a^j \partial_j \alpha_i + \nabla_i \rho^j_a \alpha_j,
\\
\anablab_a \alpha^b
&=& \rho^i_a \partial_i \alpha^b + \rho^i_c \omega_{ai}^b \alpha^c
+ C_{ac}^b \alpha^c
\nonumber \\
&=& \rho^i_a \nabla_i \alpha^b - T_{ac}^b \alpha^c,
\\
\anablab_a \beta_{b}
&=& \rho^i_a \partial_i \beta_b
- \rho^i_b \omega_{ai}^c \beta_c
- C_{ab}^c \beta_c
\nonumber \\
&=& \rho^i_a \nabla_i \beta_b + T_{ab}^c \beta_c,
\\
{}^A \rd^{\nabla}_{[a} \beta_{b]}
&=& \rho^i_a \partial_i \beta_b
- \rho^i_b \partial_i \beta_a 
- C_{ab}^c \beta_c 
\nonumber \\
&=& \rho^i_a(\partial_i \beta_b - \omega_{bi}^c \beta_c) 
- \rho^i_b(\partial_i \beta_a - \omega_{ai}^c \beta_c) 
+ T_{ab}^c \beta_c
\nonumber \\
&=& \rho^i_a \nabla_i \beta_b
- \rho^i_b \nabla_i \beta_a 
+ T_{ab}^c \beta_c,
\end{eqnarray}

Local coordinate expressions of 
a curvature, $R \in \Omega^2(M, A \otimes A^*)$,
an $A$-torsion, $T \in \Gamma(A \otimes \wedge^2 A^*)$, 
and a basic curvature, and $S \in \Omega^1(M, \wedge^2 A^* \otimes A)$, 
are given by
\beqa 
R_{ijb}^a &\equiv& 
\partial_i \omega_{bj}^a - \partial_j \omega_{bi}^a 
- \omega_{bi}^c \omega_{cj}^a + \omega_{bj}^c \omega_{ci}^a,
\label{curvatureona}
\\
T_{ab}^c &\equiv& 
- C_{ab}^c + \rho_a^i \omega_{bi}^c - \rho_b^i \omega_{ai}^c,
\label{atorsionona}
\\
S_{iab}^{c} &\equiv& 
\nabla_i T_{ab}^c + \rho_b^j R_{ija}^c - \rho_a^j R_{ijb}^c,
 \nonumber \\
&=& - \partial_i C^c_{ab} + \omega_{di}^c C_{ab}^d - \omega_{ai}^d C_{db}^c - \omega_{bi}^d C_{ad}^c
+ \rho_a^j \partial_j \omega_{bi}^c
- \rho_b^j \partial_j \omega_{ai}^c
\nonumber \\ && 
+ \partial_i \rho_a^j \omega_{bj}^c
- \partial_i \rho_b^j \omega_{aj}^c
+ \omega_{ai}^d \rho_d^j \omega_{bj} ^c
- \omega_{bi}^d \rho_d^j \omega_{aj} ^c
\label{bcurvatureona}
\eeqa
where $\nabla_i T_{ab}^c$ is
\beqa 
\nabla_i T_{ab}^c &\equiv& 
\partial_i T_{ab}^c
+ \omega_{di}^c T_{ab}^d - \omega_{ai}^d T_{db}^c 
- \omega_{bi}^d T_{ad}^c.
\eeqa
Covariant expressions of Eqs.~\eqref{LAidentity1}
and 
\eqref{LAidentity2} are 
\begin{eqnarray}
&& \rho_a^j \nabla_j \rho_b^i - \rho_b^j \nabla_j \rho_a^i 
+ \rho_c^i T_{ab}^c =0,
\label{LAidentity01}
\\
&& \rho_c^i \nabla_i T_{ab}^e - T_{ab}^d T_{cd}^e
- \rho_b^i \rho_c^j R_{ija}^e + (abc \ \mbox{cyclic}) = 0.
\label{LAidentity02}
\end{eqnarray}
The basic curvature satisfies a Bianchi type identity
${}^A \rd^{\nabla} \baS = 0$.
In local coordinate expression it is
\begin{eqnarray}
&& \rho_{[a|}^j \nabla_{j} S_{i|bc]}^d + T_{[ab|}^e S_{ie|c]}^d
- T_{[a|e}^d S_{i|bc]}^e + (\nabla_i \rho^j_{[a|}) S_{j|bc]}^{d} 
= 0.
\label{BianchiofS}
\end{eqnarray}

\newcommand{\bibit}{\sl}



\end{document}